%% file: main.tex
\documentclass[submission,copyright,creativecommons]{eptcs}
\providecommand{\event}{GCM 2022} % Name of the event you are submitting to

\usepackage{iftex}
\usepackage[framemethod=tikz]{mdframed}
\usepackage{soul}
\ifpdf
  \usepackage{underscore}         % Only needed if you use pdflatex.
  \usepackage[T1]{fontenc}        % Recommended with pdflatex
\else
  \usepackage{breakurl}           % Not needed if you use pdflatex only.
\fi

\title{Time and Space Measures for a\\ Complete Graph Computation Model}
\author{Brian Courtehoute and Detlef Plump
\institute{Department of Computer Science, University of York, York, UK}
\email{\{bc956,detlef.plump\}@york.ac.uk}
}
\def\titlerunning{Time and Space Measures for a Complete Graph Computation Model}
\def\authorrunning{B. Courtehoute \& D. Plump}

\input{macros}

\begin{document}
\maketitle

\begin{abstract}
We present a computation model based on a subclass of GP\,2 graph programs which can simulate any off-line Turing machine of space complexity $\mathrm{O}(s(n)\log s(n))$ in space $\mathrm{O} (s(n))$. The simulation only requires a quadratic time overhead. Our model shares this property with Sch\"onhage's storage modification machines and Kolmogorov-Uspenskii machines~\cite{VanEmdeBoas89a}. These machines use low-level pointer instructions whereas our GP\,2-based model uses pattern-based transformation rules and high-level control constructs.

% The abstract should briefly summarize the contents of the paper in
% 15--250 words.

% \keywords{Graph computation models  \and GP\,2 \and Complexity measures \and Space compression.}
\end{abstract}

\section{Introduction}

Sch\"onhage's storage modification machines (SMMs) and Kolmogorov-Uspenskii machines (KUMs) are graph-based computation models that do not use graph transformation rules. They have the remarkable feature of being able to simulate Turing machines using less space with only quadratic time overhead, using a uniform space measure \cite{VanEmdeBoas89a}. Although these are graph-based models, the literature on rule-based graph transformation so far seems to have ignored the work of Sch\"onhage \cite{Schonhage80a} and Kolmogorov \cite{Kolmogorov-Uspenskii63a}, and in particular the results of van Emde Boas \cite{VanEmdeBoas89a} and Luginbuhl \cite{Luginbuhl-Loui93a}.

One fundamental difference between SMMs and KUMs on the one hand, and GP\,2 on the other hand is that the inputs and outputs of SMMs and KUMs are strings, whereas GP\,2 programs compute relations on graphs. Moreover, SMMs and KUMs differ from modern graph transformation languages in that they use low-level pointer instructions instead of pattern-based rules where the programmer can specify a subgraph and how it is transformed, allowing for a more natural implementation of graph algorithms. %This is a higher-level problem-oriented approach.

An advantage of structured programming languages such as GP\,2 is the absence of ``go to'' statements which are considered to be harmful \cite{Dijkstra68a}. Furthermore, GP\,2 is based on the double-pushout approach for graph transformation, which comes with a large amount of theoretical results. We are not aware of any comparable theory for SMMs and KUMs.

% Instead of only consisting of operational constructs, it comes with a theoretical foundation such as double-pushouts, which facilitate formal proofs. Furthermore, in GP\,2 for instance, programmer-specified rules along with its control flow allow for a more natural implementation of algorithms.

The fact that the above mentioned space compression property applies to a high-level language is not obvious. Even so, we show in this paper that GP\,2 exhibits the same space compression feature. Specifically, we show that a Turing machine using $\mathrm{O}(s(n)\log s(n))$ space can be simulated in $\mathrm{O} (s(n))$ space with quadratic time overhead, where $s(n)$ is an arbitrary function. The complexity class of the simulation is a strict subset of the original class (for non-constant $s(n)$). Rather than using full GP\,2, we show that a subset of the language suffices to establish the compression result. 

Note that another way to show space compression is to efficiently simulate SMMs in general. This would also mean other properties are applicable to GP\,2, such as the real-time simulation of Turing machines shown by Luginbuhl \cite{Luginbuhl-Loui93a}. This paper's approach however comes with a concrete class of programs that show in much detail that the simulation is possible. In particular, we show \emph{how} to establish a result akin to van Emde Boas' in a rule-based graph transformation language such as GP\,2. Moreover, the van Emde Boas paper lacks a lot of technical details, which we provide in our setting.

\medskip
In Section \ref{sec:graph-transformation}, we define the subset of GP\,2 sufficient to get our results. We also show how this subset finds matches for rules efficiently.

We show how we simulate Turing machines in Section \ref{sec:simulation}. We describe the encoding of Turing machine configurations, give the simulation, provide an example, and prove correctness.

In Section \ref{sec:models} we identify a class of efficient computation models based on GP\,2, which come with time and space complexity measures. We then show that our simulation gives rise to such a model.

Finally, in Section \ref{sec:complexity}, we show the aforementioned time and space complexities of the simulation, and discuss the use of logarithmic versus uniform space measures.

\section{Rule-Based Graph Transformation}\label{sec:graph-transformation}

Throughout this paper, we use a subset of GP\,2 called SGP\,2 (Small GP\,2, defined in Subsection \ref{subsec:small-gp2}), which suffices to establish the simulation and compression result. This subset uses simpler host graphs. Specifically labels are limited to being either an integer or one of three characters, and lists of these are not allowed. Rules are simpler as well. They use neither label expressions with variables, nor application conditions.

In Subsection \ref{subsec:matching}, we analyse the complexity of rule matching by showing a new variant of a Theorem that allows for constant-time rule matching.

\subsection{Small GP\,2}\label{subsec:small-gp2}

SGP\,2 programs transform input graphs into output graphs, where graphs are directed and may contain parallel edges and loops. Both nodes and edges are labelled.

\begin{definition}[Graph]
\label{def:graph}
\normalfont
Let $\L$ be the set of labels defined by the grammar in Figure \ref{fig:label-grammar}. A \emph{graph}\/ over $\L$ is a system $\tuple{V,E,P,s,t,l,m}$, where $V$ and $E$ are finite sets of nodes (or vertices) and edges, $P\subseteq V$ is the set of roots, $s\colon E \to V$\/ and $t\colon E \to V$\/ are source and target functions for edges, $l\colon V \to \L$ is the partial node labelling function and $m\colon E \to \L$ is the (total) edge labelling function.
\end{definition}

% To speed up matching, SGP\,2 supports \emph{rooted} graph transformation where graphs in rules and host graphs are equipped with so-called root nodes. Roots in rules must match roots in the host graph so that matches are restricted to the neighbourhood of the host graph's roots.

A graph is \emph{totally labelled}\/ if $l$ is a total function. Unlabelled nodes will only occur in the interfaces of rules and are used to relabel nodes. There is no need to relabel edges as they can always be deleted and reinserted with different labels.

\input{Figures/gp2/label-grammar}

Both nodes and edges are labelled with either an integer, with the special characters \ttt{"L"}, \ttt{"R"}, \ttt{"I"}, or with the constant \ttt{empty}. By convention, we draw items labelled with \ttt{empty} in pictures without labels. Figure \ref{fig:label-grammar} defines the labels that may occur in both rules and host graphs. Integers within a label are represented in base $10$. In this paper however, we only use digits up to $2$ in labels.

We call the graphs to which rules are applied \emph{host graphs}, they consist of all totally labelled graphs.

Nodes and edges can be \emph{marked} red, green or blue. In addition, nodes can be marked grey and edges can be dashed. For example, rules in \ttt{CacheNext} in Figure \ref{fig:encode} contains red and unmarked nodes and blue edges. Marks are convenient, among other things, to record visited items during a graph traversal and to differentiate between similar structures.

The principal SGP\,2 programming construct used in this paper are graph transformation rules according to the double-pushout approach with injective matching. For example, the rules in \ttt{CacheNext} in Figure \ref{fig:encode} consist of a left-hand graph and a right-hand graph which are specified graphically. The small numbers attached to nodes are identifiers, all other black text in the graphs are labels.

\begin{definition}[Rule]
\label{def:rule}
\normalfont
A \emph{rule} $\tuple{L \gets K \to R}$ consists of totally labelled graphs $L$\/ and $R$, a graph $K$\/ consisting of unlabelled nodes only, where $V_K \subseteq V_L$\/ and $V_K \subseteq V_R$.
\end{definition}

When a rule is declared, such as in \ttt{CacheNext} in Figure \ref{fig:encode}, graph $K$\/ is implicitly represented by the node identifiers in $L$\/ and $R$\/ (which much coincide). Nodes without identifiers in $L$\/ are to be deleted and nodes without identifiers in $R$\/ are to be created. Roots are represented by double circles.

\begin{definition}[Graph Morphism]
\label{def:premorphism}
\normalfont
 Given graphs $L$\/ and $G$, a \emph{graph morphism} $g\colon L \to G$\/ consists of two functions $g_V\colon V_L \to V_G$ and $g_E\colon E_L \to E_G$\/ that preserve sources, targets and labels, that is, $s_G \circ g_E = g_V \circ s_L$, $t_G \circ g_E = g_V \circ t_L$, $l_G \circ g_V = l_L$ and $m_G \circ g_E = m_L$. Moreover, we require that roots are preserved and reflected, that is, $g_V(P_L) \subseteq P_G$ and $g_V^{-1}(P_G) \subseteq P_L$.
\end{definition}

\begin{definition}[Rule Application]
\label{def:rule-application}
\normalfont
A rule $r = \tuple{L \gets K \to R}$ is applied to a host graph $G$ as follows:
\begin{enumerate}
\item Find an injective graph morphism $g\colon L \to G$\/ that satisfies the \emph{dangling condition}: no node in $g_V(V_L - V_K)$ is incident to an edge in $E_G - g_E(E_L)$.
\item Construct a host graph $H$\/ from $G$\/ as follows:
\begin{samepage}
\begin{enumerate}
\item Remove all edges in $g_E(E_L)$ and all nodes in $g_V(V_L) - g_V(V_K)$, obtaining a subgraph $D$ of $G$.
\item $V_H = V_D + (V_R - V_K)$ and $E_H = E_D + E_R$.
\item For each edge $e \in E_D$, $s_H(e) = s_D(e)$. For each edge $e \in E_R$, $s_H(e) = s_R(e)$ if $s_R(e) \in V_R-V_K$, otherwise $s_H(e) = g_V(s_R(e))$. Targets are defined analogously.
\item For each node $v \in V_D - g_V(V_K)$, $l_H(v) = l_D(v)$. For each node $v \in g_V(V_K)$ with $g_V(\bar{v}) = v$, $l_H(v) = l_{R}(\bar{v})$. For each node $v \in V_R - V_K$, $l_H(v) = l_{R}(v)$.
\item For each edge $e \in E_D$, $m_H(e) = m_D(e)$. For each edge $e \in E_R$, $m_H(e) = m_{R}(e)$.
\item $P_H = ( P_G - g_V(V_L) ) \cup P_R$.
\end{enumerate}
\end{samepage}
\end{enumerate}
We write $G \dder_{r,g} H$\/ to express that $H$\/ results from $G$ by applying $r$ with match $g$.
\end{definition}

\input{Figures/gp2/program-syntax}

SGP\,2 programs follow the syntax described in Figure \ref{fig:command-syntax}. We omit the the definition of RuleDecl because we do not use textual rule declaration in this paper. We declare rules graphically so they are easier to read. An example of a graphical rule declaration can be seen in Figure \ref{fig:simulation}. The structural operational semantics of programs is defined in Figure \ref{fig:core-semantics}, where inference rules inductively define a small-step transition relation $\to$ on \emph{configurations}. In the setting of SGP\,2, a configuration is either a command sequence together with a host graph, just a host graph or the special element fail:
\[ \to \;\; \subseteq \; (\text{ComSeq} \times \G) \times 
      ((\text{ComSeq} \times \G) \cup \G \cup \{\failrm\}) \]
where $\G$ is the set of SGP\,2 host graphs. Configurations in $\text{ComSeq} \times \G$, given by a rest program and a host graph, represent states of unfinished computations while graphs in $\G$ are final states or \emph{results} of computations. The element fail represents a failure state. A configuration $\gamma$ is said to be \emph{terminal} if there is no configuration $\delta$ such that $\gamma \to \delta$.

% An example of a SGP\,2 program can be found in Figure \ref{fig:example-prog}.

% \input{Figures/gp2/example}

Figure \ref{fig:core-semantics} shows the inference rules for the core commands of SGP\,2. We use an older version of the semantics because it is simpler, and equivalent to the current one for terminating programs (which is all we care about for time complexity analysis) \cite{Courtehoute-Plump21a}. The rules contain meta-variables for command sequences and graphs, where $R$ stands for a call of a rule set or of a rule, $C,P,P',Q$ stand for command sequences, and $G,H$ stand for host graphs. Branching statements without \ttt{then}/\ttt{else} clauses are omitted, because they can be defined with the other commands.

The transitive and reflexive-transitive closures of $\to$ are written $\to^+$ and $\to^*$, respectively. We write $G \dder_R H$ if $H$ results from host graph $G$ by applying the rule set $R$, while $G \not\dder_R$ means that there is no graph $H$ such that $G \dder_R H$ (application of $R$ fails).

\input{Figures/gp2/core-semantics}

\subsection{Time Complexity of Rule Matching}\label{subsec:matching}

When analysing the time and space complexity of programs, we assume that these are fixed. This is customary in algorithm analysis where programs are fixed and running time or space is measured in terms of input size \cite{Aho-Hopcroft-Ullman74a,Skiena20a}. In our setting, the input size is the \emph{size} of a host graph, which we define to be the total number of nodes and edges.
When we match a rule graph $L$ into a host graph $G$, $L$ is constant since it is given by the program, but $G$ is not since it depends on the input.

Since unrestricted matching is very time-consuming we use rooted rules. They contain special nodes called \emph{roots} that can match with other roots in the host graph in constant time, allowing rule matching to happen locally.

In order to achieve this efficient rule matching, we use a variant of the fast matching theorem first introduced in \cite{Bak-Plump12a}. The trade-off with the original is that we can have a more general host graph, i.e. we allow the indegree to be unbounded, but there are more restrictions on rules, namely there needs to be a directed path from some root node to every node in the left-hand side.

To prove that matching can be implemented in constant time, we give Algorithm \ref{algo:matching}, which matches a graph $L$ into a graph $G$. It is identical to the one given in \cite{Bak15a}, except for the following two details. First, we consider morphisms instead of pre-morphisms (which ignore labels) since label expressions with variables or lists are not part of SGP\,2. Second, we adjust the definition of edge enumerations to take into account edge direction. Because these are the only changes from the algorithm in \cite{Bak15a}, a correctness proof would be very similar and is hence omitted.

The definition of edge enumeration is very similar to the one given in \cite{Bak15a}, except that every time an edge is required to be incident to a node, we require that node to be the source of the edge. This is to make sure edges are matched via their source node. Since we use host graphs of bounded outdegree only, this ensures the number of matches is bounded.

\begin{definition}[Edge Enumeration]
    Given a graph $L$ and a node $p \in P_L$ (the set of roots of $L$), an \emph{edge enumeration} $e_1,\dots,e_n$ for $p$ is a list of edges such that $s(e_1)=p$, and for each $i\geq 2$, $s(e_i)$ is the source or target of some edge in $e_1,\dots,e_{i-1}$.
\end{definition}

Note that the source of each edge in an enumeration of $p$ is reachable from $p$ via a directed path. Furthermore, the union of edge enumerations over all roots of $L$ only cover edges reachable from a root via a directed path. If we restrict the left-hand side of rules to only have nodes reachable from a root, the source of each edge must be reachable from a root, and so must the edge itself. Matching algorithms for other rules are out of the scope of this paper and can be found in \cite{Bak15a}.

\begin{definition}[Fast Rule]
    A rule $L \dder R$ is \emph{fast} if each node in $L$ is reachable from some root via a directed path.
\end{definition}

The other requirements for a rule to be fast from \cite{Bak-Plump12a} are omitted since they concern variable labels and application conditions, which do not exist in SGP\,2.

\begin{definition}[Partial Morphism]
    A graph morphism $g\colon L' \to G$ is a \emph{partial morphism} $g\colon L \xrightarrow{par} G$ if $L' \subseteq L$. We say that $g$ is \emph{undefined} for the items in $L-L'$ and call $L'$ the \emph{domain} of $g$, denoted by $\mathrm{Dom}(g)$.  
\end{definition}

Consider partial morphisms $f,g\colon L\xrightarrow{par}G$ such that $f(x) = g(x)$ for all items $x$ in $\mrm{Dom}(g)$. We say that $f$ \emph{extends} $g$ by a node $v$ if $\mrm{Dom}(f)_V = \mrm{Dom}(g)_V \cup \{v\}$ and $\mrm{Dom}(f)_E = \mrm{Dom}(g)_E$. We say that $f$ extends $g$ by an edge $e$ if $\mrm{Dom}(f)_E = \mrm{Dom}(g)_E \cup \{e\}$ and $\mrm{Dom}(f)_V = \mrm{Dom}(g)_V \cup \{s_G(e),t_G(e)\}$.

\begin{algorithm}
\DontPrintSemicolon % Some LaTeX compilers require you to use \dontprintsemicolon instead
\KwIn{Graphs $L$ and $G$; and for each $p \in P_L$, an edge enumeration $e_{p_1},\dots,e_{p_n}$}
\KwOut{The set $A$ of all injective graph morphisms $L \rightarrow G$ that are root-preserving and root-reflecting}
    A $\gets \{h:L\xrightarrow{par} G \mid \mathrm{Dom}(h) = \emptyset\}$\;
    \While{there is an untagged root $p \in P_L$}{
        \begin{tabular}{ p{5mm}  p{1000mm} }
        $A_0 \gets$
        & $ \{h\colon L\xrightarrow{par} G \mid \text{$h$ is injective, root-preserving and root-reflecting, and there exists }$
        \\
        & $\text{$h'\in A$ such that $h$ extends $h'$ by p}\}$
        \end{tabular}
        \;\vspace{-5mm}
        tag $p$\;
        \For{$i=1$ to $n$}{
            \begin{tabular}{ p{5mm}  p{1000mm} }
            $A_{p_i} \gets$
            & $ \{h\colon L\xrightarrow{par} G \mid \text{$h$ is injective, root-preserving and root-reflecting, and there exists }$
            \\
            & $\text{$h' \in A_{p_i-1}$ such that $h$ extends $h'$ by $e_{p_i}$}\}$
            \end{tabular}
            \;\vspace{-5mm}
            \lIf{$s(e_{p_i})\in P_L$}{tag $s(e_{p_i})$}
            \lIf{$t(e_{p_i})\in P_L$}{tag $t(e_{p_i})$}
        }
        $A \gets A_{p_n}$
    }
    \Return{$A$}
\caption{A graph matching algorithm}
\label{algo:matching}
\end{algorithm}

In order to analyse the complexity of Algorithm \ref{algo:matching} below, we need to assume that the extending a partial morphism by a single node or edge takes constant time. Constant-time integer comparison is a common assumption in algorithm analysis. Only in applications that require computing with large integers is a more sensitive model needed \cite{Aho-Hopcroft-Ullman74a}.

\begin{theorem}[Fast Rule Matching]\label{thm:matching}
Matching can be implemented to run in constant time for fast rules, using host graphs with a bounded outdegree that contain a bounded number of roots.
\end{theorem}
\begin{proof}
% \begin{mdframed}[hidealllines=true,backgroundcolor=yellow,innerleftmargin=3pt,innerrightmargin=3pt,leftmargin=-3pt,rightmargin=-3pt]
    We argue that Algorithm \ref{algo:matching} implements matching as described by the theorem.
% \end{mdframed}    
    
    For this proof, we show that the number of times a morphism is updated with a single node or edge is bounded. Note that $L$ has a bounded number of nodes and edges. So subsets of nodes or edges of $L$ such as $P_L$ and edge enumerations in $L$ are also bounded. The while loop has constantly many iterations because it terminates once every node in the bounded set $P_L$ is tagged, which happens at least once in every iteration. The for loop has constantly many iterations because it iterates over an edge enumeration.
    
    In line $3$, a partial morphism is added to $A_0$ for each match of the root $p$ in $G$. Since we use host graphs with a bounded number of roots, the number of these matches is constant as well. Each of these morphisms extends the empty morphism by one node, so there are constantly many single-node morphism updates.
    
    In line $8$, we extend morphisms by an edge $e_{p_i}$, whose source is already in the extended morphism by definition of edge enumerations. Since $G$ has bounded outdegree, the number of matches for $e_{p_i}$ is bounded, and each morphism can only be extended in constantly many ways. The fact that there can only be constantly many morphisms can be shown by induction on $A_0,A_1,A_2,\dots$. By the previous paragraph, $A_0$ has bounded size. For $k\geq0$, if $A_k$ has bounded size, then so does $A_{k+1}$ since each of the morphisms in $A_k$ can only be extended in constantly many ways. So there are constantly many single-edge morphism updates. Additionally, for each such update, the target of the edge may need to be added to the definition of the morphism. Hence there are also constantly many single-node morphism updates.

\end{proof}

\section{Simulating Turing Machines}\label{sec:simulation}

First, we define what Turing machines we are simulating in Subsection \ref{subsec:off-line-tm}. We then outline the basic idea of how we achieve space compression in Subsection \ref{subsec:simulation-idea}. In Subsection \ref{subsec:configuration-graphs}, we describe exactly how Turing machine configurations are represented as graphs. Next, we give and describe the class of programs that simulate Turing machines in Subsection \ref{subsec:simulation}. We then provide an example of a Turing machine simulation in Subsection \ref{subsec:examples}. Finally, we show the correctness of the simulation in Subsection~\ref{subsec:correctness}.

\subsection{Off-Line Turing Machines}\label{subsec:off-line-tm}

In this paper, we consider deterministic off-line Turing machines in order to exhibit that space compression can happen with sublinear space complexities.

\begin{figure}[h]
    \centering
\usetikzlibrary{positioning,calc}
\begin{tikzpicture}[every node/.style={block},
        block/.style={minimum height=1.5em,outer sep=0pt,draw,rectangle,node distance=0pt}]
    \node (0) {$1$};
    \node (1) [right=of 0] {$1$};
    \node (2) [right=of 1] {$0$};
    \node (3) [right=of 2] {$1$};
    \node (4) [right=of 3] {\ttt{\sdots}};
    \node (5) [right=of 4] {$0$};
    \node (6) [right=.35 of 5,draw=none] {input tape};
    \node (H) [below = 0.4cm of 0,circle,thick,inner sep=1pt] {$q_0$};
       \draw[-latex] (H) -- (0);
       
    % \draw 
    % % (0.north west) -- ++(-.5cm,0) (0.south west) -- ++ (-.5cm,0) 
    %       (5.north east) -- ++(.5cm,0) (5.south east) -- ++ (.5cm,0);

    \node (10) [below = 0.4cm of H] {$2$};
    \node (11) [right=of 10] {$2$};
    \node (12) [right=of 11] {$2$};
    \node (13) [right=of 12] {$2$};
    \node (14) [right=of 13] {$2$};
    \node (15) [right=of 14,draw=none] {\ttt{\sdots}};
    \node (16) [right=.35 of 15,draw=none] {working tape};
    % \draw (10.north west) -- ++(-.5cm,0) (10.south west) -- ++ (-.5cm,0); 
    \draw (14.north east) -- ++(.5cm,0) (14.south east) -- ++ (.5cm,0);
    \draw[-latex] (H) -- (10);
\end{tikzpicture}
    \caption{Initial configuration of an off-line Turing machine}
    \label{fig:input-tape}
\end{figure}
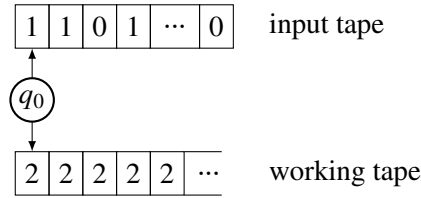

The input tape is finite and uses a binary encoding, i.e. the input alphabet is  $\Sigma = \{0,1\}$. It is read-only (symbols cannot be modified) with a tape head pointing towards the tape square being read. The input tape contains the input of the Turing machine, and its length is the length of the input.

Off-line Turing machines additionally have a working tape, a one-way infinite tape that can be read and written on (symbols can be modified). The tape heads on both tapes move simultaneously, but independently. We allow for an additional symbol $2$ as a blank symbol or separator, i.e. the working tape alphabet is $\Gamma=\{0,1,2\}$. We use $2$ instead of another symbol because it is useful for the working tape to be encoded in base $3$ for the purposes of our simulation. Initially, the working tape head contains only blank symbols, i.e. the symbol $2$. We define \emph{working tape contents} to be the section of the working tape starting at the beginning of the tape, and ending with the final nonblank (not a $2$) tape square that is followed by only blanks (only $2$'s).

We consider the finite state set $Q$ to consist of integers. There is an initial state $q_0$, an accepting state $h_a$, and a transition function $\delta \colon Q \times \Sigma \times \Gamma \to Q \times \Gamma \times \{\textrm{L},\textrm{R},\textrm{S}\} \times \{\textrm{L},\textrm{R},\textrm{S}\}$. Consider the transition $\delta(q,a,b) = \tuple{q',b',D_1,D_2}$. This means in state $q$, with $a$ on the input tape, and $b$ on the working tape, the machine goes into state $q'$, replaces the working tape symbol with $b'$, moves the input tape head in direction $D_1$, and moves the working tape head in direction $D_2$. Transitions and computations are defined as usual.

The time complexity of a Turing machine is the function which, to an integer $n$, associates the maximum number of transitions (or length of computation) starting from an input of size $n$. The space complexity is the function which associates to $n$ the maximum number of used working tape squares for an input of size $n$.

\subsection{The Basic Idea of the Simulation}\label{subsec:simulation-idea}

\medskip
We simulate a Turing machine $M$ with an SGP\,2 program called \ttt{Sim}$(M)$, as specified in Figures \ref{fig:simulation} to \ref{fig:restart}. Input symbols, tape symbols, and state names are encoded as integers. The compression is achieved by representing blocks of Turing machine tape squares as single nodes. This idea is shared with van Emde Boas' representation \cite{VanEmdeBoas89a}. By representing graphs in SGP\,2, we can have an edge that provides a direct link to a distant list element, which is impossible in a Turing machine.

\input{Figures/Examples/working-tape}

The space compression comes from how we represent the working tape, which is illustrated in Figure~\ref{fig:working-tape} (note that the text not contained within nodes or squares is explanatory). The working tape is divided into blocks of size $c=4$. Each block can be interpreted as a $4$-digit number in base $3$. Block $1$ for instance contains the ternary number $1022$. In base $10$, that is $1\cdot 3^3+0\cdot 3^2+2\cdot3^1+2\cdot3^0=27+6+2=35$. Hence we say the content of block $1$ is $35$. The graph CACHE represents the block currently containing the working tape head, which is block $1$ in this case. We call this block the \emph{active block}. CACHE is a doubly-linked list whose $4$ nodes are labelled with the symbol of the corresponding tape square. The graph BLOCKSET is a less direct representation of the working tape. Each of its nodes represents a block of working tape squares instead of a single square. Instead of storing the content of a block as a spacious label, we store it as a single dashed edge. (This is the crucial idea making compression possible.) Block $79$ for instance has content $3$, so there is a dashed edge from  node ``block $79$'' to  node ``block $3$''. The content of a block is converted into the index of a node in BLOCKSET (seen as a doubly-linked list). We call the nodes of BLOCKSET \emph{block nodes}.

In general, if block $i$ has content $j$, there is dashed edge in BLOCKSET from  node ``block $i$'' to  node ``block $j$''. The only exception is the node representing the active block, in this case node ``block $1$'', which has a dashed edge pointing towards node ``block $0$''. This is by convention for the following reason. The content of ``block $1$'' is stored in CACHE, so storing it in BLOCKSET as well would be redundant. So instead of using the dashed edge to store the content of ``block $1$'', the dashed edge is directed to the leftmost node (``block $0$'') by convention. Since that dashed edge now holds no information, when labels in CACHE are changed, the dashed edge does not need to be updated. Once the tape head moves to a different block, and the active block changes, that dashed edge is redirected to represent to content of its block again, and the outgoing dashed edge of the new active block node is redirected to node ``block $0$''.

The only part of the tape we need to represent is the working tape contents, since only blanks follow. For a given working tape, let $b$ be the minimum number of blocks that contain the working tape contents. So we only need $b$ nodes in BLOCKSET. Remember $c=4$ is the number of tape squares per block. Since we are converting the content of a block into a list index in BLOCKSET, the list needs to be long enough to have that index. In other words, the number of nodes $b$ in BLOCKSET needs to be greater than the largest possible block content. A $4$-digit ternary number can have $3^c=3^4=81$ different values, so we need $b \geq 3^c = 81$. Since our aim is compressing space, we want to use the lowest number of nodes, so we pick $b=3^c=81$. If the tape contents exceed the size $b$ blocks, we extend our representation by incrementing $c$, and updating $b=3^c$. These new values become $c=5$ and $b=243$.

After incrementation, the number of nodes in CACHE is $c=5$, and the number of edges $2\cdot(c-1)=8$, so $13$ in total. The number of nodes in BLOCKSET is $b=243$, and the number of edges $b+2\cdot(b-1)=727$, so $970$ in total. However the number of squares on the working tape is $b \cdot c = 19683$, which is significantly larger. The graph representation uses less space for $c\geq 5$. Let us describe this behaviour asymptotically. We have $c = \log_3 b$. The number of outgoing edges of a node is bounded by $3$, hence the number of nodes and edges is bounded by $4(b+c)$, which is $\mathrm{O}(b + \log_3 b) = \mathrm{O}(b)$. The number of working tape squares on the other hand is $\mathrm{O}(b \log_3 b)$.

% Note that the number of outgoing edges for each node in BLOCKSET and CACHE is bounded by $3$. So the number of nodes and edges is bounded by $4\cdot(b+c)=4\cdot(81+4)=340$.

% We encode the working tape of a Turing machine by dividing it into $b$ blocks of equally many tape squares, which we represent as a list BLOCKSET of \emph{block nodes}. The tape symbols in a block form the digits of a ternary number $n$. That number is stored as a dashed edge leaving the block node, pointing to the block node of index $n$. For example, in Figure \ref{fig:graph-representation}, all block nodes point to the block node of index $0$, meaning the tape symbols of each block are all $0$. The symbols of the first block, which is currently the \emph{active block}, are explicitly represented in the list CACHE, which consists of as many nodes as there are squares in a block. The active block is the one containing the working tape head.

% Hence the number of represented tape squares is $b \cdot \log_3 b$, the number of blocks times the number tape squares per block. However, we represent these using the nodes of BLOCKSET and CACHE, of which there are only $b + \log_3 b$. In addition, each of these nodes has at most $3$ outgoing edges, and thus the total graph size is still $\mathrm{O}(b + \log_3 b)$, which is the same as $\mathrm{O}(b)$.

\subsection{Turing Machine Configurations as Graphs}\label{subsec:configuration-graphs}

Figure \ref{fig:input-tape} shows the initial configuration of a Turing machine with its input and working tapes and tape heads, as well as the initial state $q_0$. The corresponding graph can be seen in Figure \ref{fig:graph-representation}.

\input{Figures/Examples/example}

Let $\mathrm{O}(s(n)\log s(n))$ be the space complexity of a Turing machine, where $n$ is the size of the input tape, and $s$ some function. For the rest of this paper, we omit the base of the logarithm since $\mathrm{O}(\log_3 n) = \mathrm{O}(\log_c n)$ for any constant $c>1$. If we choose the number of blocks to be $b=\mathrm{O}(s(n))$, we can represent $\mathrm{O}(s(n)\log s(n))$ tape squares in space $\mathrm{O}(s(n))$, as outlined by the previous subsection. This will be proven in Theorem \ref{thm:space-complexity}. 

The graph representation consists of the \emph{central node}, labelled by the initial state $q_0$ as represented by some integer, and the subgraphs INPUT, BLOCKSET, and CACHE. We represent `left' and `right' positioning by blue and red edges respectively. These edges will not be modified by the program. The unmarked and unlabelled green edges mark the positions of the tape heads. Dashed edges serve to encode the state of BLOCKSET and how it relates to CACHE, as described in the previous subsection. The dashed edge from the central node points towards the block node whose block currently contains the working tape head (the active block node). By convention, there is a dashed edge from the active block node to the first block node, as stated in the previous subsection. Hence there is always a path of dashed edges from the central node to the active block node to the first block node.

The central node is a root that is never moved. Parts of the graph such as tape head positions and tape ends can be accessed efficiently via the central node's outgoing edges. The only other time roots are used in the simulation is when extending the tape. We use roots to keep track of the end of the tape, and later to traverse the tape. For a more in-depth explanation of how the roots move, see Subsection \ref{subsec:simulation}.

INPUT represents the input tape of the Turing machine, and the position of the tape head is represented by an unmarked green edge from the central node. The green edge labelled \ttt{"I"} always points to the leftmost square of the input tape. The working tape is represented by BLOCKSET and CACHE. These subgraphs are implemented as doubly linked lists in order to enable fast rule matching.

Each node in BLOCKSET represents a block of $c$ nodes of the working tape. The blue edge from the central node points towards the leftmost node, and the dashed edge towards the block that contains the position of the tape head.

CACHE is the block of the working tape that contains the tape head. Its position within BLOCKSET is marked with a dashed edge from the central node. The red edge from the central node points towards the rightmost node of the cache, and the unmarked edge towards the current position of the tape head. The node labels represent the content of each square.

The operations of the Turing machine happen within CACHE. If the tape head moves out of bounds of CACHE, we encode the content of CACHE within BLOCKSET, move on to another block, and decode it into CACHE.

    Initially, the tape heads are on the first square of their tapes, and the working tape is blank (i.e. filled with $2$'s). Remember CACHE has $c$ nodes and BLOCKSET $3^c$. We start simulation with $c=2$ because it is the smallest nontrivial value for $c$. If $c=1$, then each block would represent a single tape square, and no space would be gained. If we run out of nodes in BLOCKSET, $c$ is incremented and BLOCKSET adjusted accordingly.

\subsection{The Simulation \ttt{Sim}$(M)$}\label{subsec:simulation}

In this subsection, we assume we have a Turing machine $M$ and describe \ttt{Sim}$(M)$, the SGP\,2 program that simulates $M$. The program \ttt{Sim}$(M)$ takes inputs that consist of graphs such as INPUT together with the central node and green edges as represented in Figure \ref{fig:graph-representation}. The overall behaviour of the program is that it starts the simulation with a small BLOCKSET and CACHE. If we run out of tape squares, we reset to the initial configuration, extend BLOCKSET and CACHE, and restart the simulation.

Figure \ref{fig:simulation} contains the main control sequence of $\ttt{Sim}(M)$. First the rule \ttt{setup} is called. It matches the root and constructs CACHE with two nodes and BLOCKSET with nine nodes, as seen in Figure \ref{fig:graph-representation}. We omit the definition of that rule since it is straightforward.

Next we have the loop \ttt{Simulate!}, which means the procedure \ttt{Simulate} is applied until no longer possible. The loop body first calls \ttt{Transitions} (Figure \ref{fig:transitions}), a procedure that applies the transition function to the current state of $M$. If \ttt{Transitions} cannot apply the transition function, the procedure results in failure, and the loop terminates. Next, \ttt{try MoveLeft} is called, which means we attempt to apply \ttt{MoveLeft} and if it fails, we skip this instruction. The rules in \ttt{MoveLeft} detect a label \ttt{"L"} on the unmarked edge adjacent to the central node, which is created in \ttt{Transitions} if the working tape head needs to move to the left. It then executes that move in CACHE. Analogously, \ttt{try MoveRight} attempts to move the working tape head to the right. These rule sets (which are nondeterministic calls of the rules they contain) will fail however if the tape head moves out of bounds of CACHE, which is detected by \ttt{Left} or \ttt{Right} since the continued presence of the \ttt{"L"} or \ttt{"R"} label indicates that tape head movement still needs to be done. In that case, \ttt{PrevBlock} or \ttt{NextBlock} (Figure \ref{fig:block}) are called in order to move towards the relevant block.

\input{Figures/Simulation/simulation}
\input{Figures/Simulation/transitions}
\input{Figures/Block/block}
\input{Figures/Encode/encode}
\input{Figures/Decode/decode}
\input{Figures/FlagOps/restart}

Figure \ref{fig:transitions} contains \ttt{Transitions}, the rule set that encodes the transition functions of $M$. For each entry in the transition table, there is a corresponding rule in \ttt{Transitions}. The rules are divided into nine categories \ttt{transitionXY}, where \ttt{X} represents the movement of the input tape head, and \ttt{Y} that of the working tape head. Each rule implements the change of labels and the movement of the input tape head directly. For the movement of the working tape head, a label is left behind on the unmarked edge, and the actual movement is handled by \ttt{Simulate} in Figure \ref{fig:simulation}.

In Figure \ref{fig:block}, we show the procedures \ttt{NextBlock} and \ttt{PrevBlock}, which handle movement to another block. We call the node pointed at by a dashed edge from the central node the \emph{active block node}. The procedure \ttt{Encode} (Figure \ref{fig:encode}) saves the information from CACHE into BLOCKSET. This process is further described in the example in Subsection \ref{subsec:examples}. The procedure \ttt{CacheInc} increments CACHE as a ternary number. First the rightmost digit is turned into a red root. Then the digit is attempted to be incremented with the rule set \ttt{Inc}. If the digit is a \ttt{0} or a \ttt{1}, incrementation succeeds, the process is marked as finished by turning the red root into an unmarked node, and \ttt{CacheInc} terminates. If the digit is a \ttt{2} however, \ttt{Inc} fails, the \ttt{2} is turned into a \ttt{0}, and the red root is moved along to the next digit which is attempted to be incremented. If \ttt{Increment!} ends without \ttt{Inc} ever succeeding, it means that all digits were \ttt{2}, and that the red root is still present. In that case, \ttt{Reset!} turns all digits into \ttt{0}.

The procedure \ttt{Decode} in Figure \ref{fig:decode} decodes the active block in BLOCKSET and stores the information in CACHE. It is analogous to \ttt{Encode} (note that some of \ttt{Decode}'s rules and procedures appear in Figure \ref{fig:encode}). Similarly to \ttt{CacheInc}, \ttt{CacheDec} decrements CACHE as a ternary number if that number is not $0$, and keeps it as $0$ if it already is.

Figure \ref{fig:restart} shows the procedure \ttt{Restart} which, when the machine runs out of tape, resets the simulation and extends the tape. CACHE and hence the size of a block is extended by one square, and the size of BLOCKSET is tripled. The procedure \ttt{RewindTapes} rewinds the tape heads to the beginning. Next, \ttt{ResetCache} resets the content of CACHE back to $2$'s (the blank symbol). Then, \ttt{ResetBlockset} extends BLOCKSET and directs dashed edges to their initial state. In rules \ttt{binit} and \ttt{copy}, an unlabelled blue root traverses the existing BLOCKSET, while blue roots with labels \ttt{1} and \ttt{2} traverse two copies of that BLOCKSET as they are being created. In the meantime, \ttt{Undirect} deletes dashed edges that are outgoing from block nodes. The rule \ttt{glue} appends the copies to the original BLOCKSET, tripling its length. The blue nodes are unrooted except for the final node, which remains a blue root labelled \ttt{2}. The penultimate node becomes an unmarked root, which is used in \ttt{direct} to traverse BLOCKSET from right to left, creating dashed edges where needed. The rule \ttt{unroot} then removes the blue and unmarked roots.

\subsection{Example}\label{subsec:examples}
 
\begin{figure}[h!]
     \centering
     \begin{subfigure}[b]{0.3\textwidth}
         \centering
         \includegraphics[width=\textwidth]{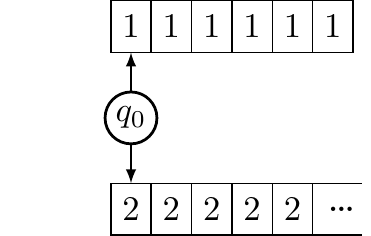}
         \caption{Turing machine}
         \label{fig:exina}
     \end{subfigure}
     \hfill
     \begin{subfigure}[b]{0.69\textwidth}
         \centering
         \includegraphics[width=\textwidth]{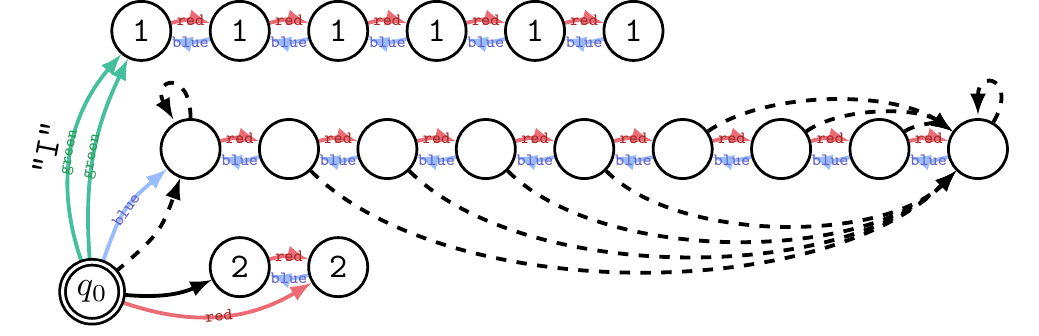}
         \caption{Simulation}
         \label{fig:exinb}
     \end{subfigure}
        \caption{Initial state of the example machine}
        \label{fig:exin}
\end{figure}

\begin{figure}[h!]
     \centering
     \begin{subfigure}[b]{0.45\textwidth}
         \centering
         \includegraphics[width=\textwidth]{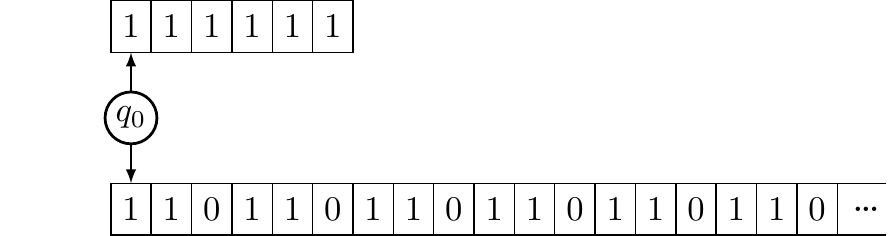}
         \caption{Turing machine}
         \label{fig:exouta}
     \end{subfigure}
     \hfill
     \begin{subfigure}[b]{0.54\textwidth}
         \centering
         \includegraphics[width=\textwidth]{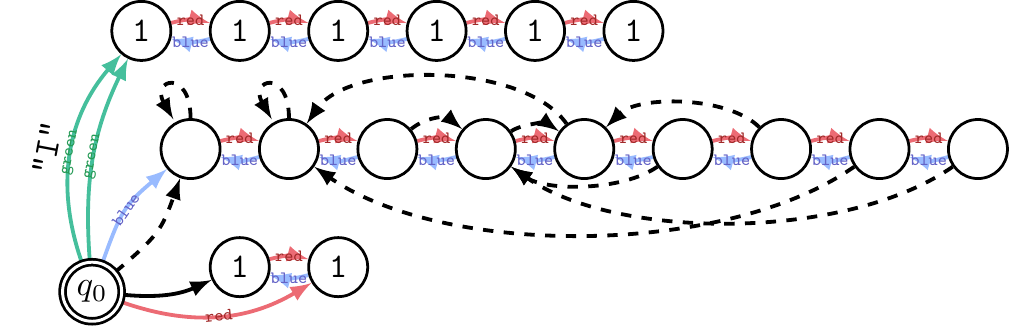}
         \caption{Simulation}
         \label{fig:exoutb}
     \end{subfigure}
        \caption{Final state of the example machine}
        \label{fig:exout}
\end{figure}

In this subsection, we give an example of a Turing machine simulation, and show how we move from one block to another. Consider a Turing machine that takes as input the number $n$ represented in unary, and writes $n$ in binary on its working tape $n$ times. It is reasonable to assume the machine has a space complexity of $\mathrm{O}(n \log n)$. If $n=6$, the machine uses $18$ squares, which are filled with $6$ copies of the string \ttt{110} ($6$ in binary). A CACHE size of $2$ and a BLOCKSET size of $9$ are enough to represent $2 \cdot 9 = 18$ tape squares. Their representation in the simulation has only $2+9=11$ nodes, which is $\mathrm{O}(n + \log n) = \mathrm{O}(n)$. The initial state of the machine on input $6$ is shown in Figure \ref{fig:exin}, and the final one in Figure \ref{fig:exout}.

Initially, the working tape is blank (all $2$'s). Hence most blocks consist of two $2$'s, i.e. their content is $8$ ($22$ in ternary is $8$ in base $10$). So most block nodes point towards the block node of index $8$ (final block node). The exception is the first node, which points to itself. This is because it represents the block that contains the tape head, which is set to index $0$ by convention. In the final state, the working tape is represented in the same manner. Note that even though the working tape content looks binary, $2$'s (blank symbols) could be present, and hence tape blocks are considered a ternary number. For instance consider the third block (of size $2$). It contains \ttt{10}, so the third bock node points towards the block node of index $3$ (fourth block node).

Let us sketch the behaviour of the machine. First, the input is copied in unary onto the working tape for use as a counter. Then, a binary number to the right of the counter is incremented while traversing the input. The previous step is repeated while decrementing the counter until it reaches $0$. Tape contents need to be shifted. The symbol $2$ can be used as a separator.

% \begin{example}[Changing blocks]
\medskip
Now let us show what happens when the block has to be changed.
Consider the situation where \ttt{0100} are the first $4$ squares of the working tape and the tape head is on the third square. Let the next transition write symbol \ttt{1} and move the tape head to the left, and change state $q_i$ to state $q_j$.

\begin{figure}[h!]
     \centering
     \begin{subfigure}[b]{0.32\textwidth}
         \centering
         \includegraphics[width=\textwidth]{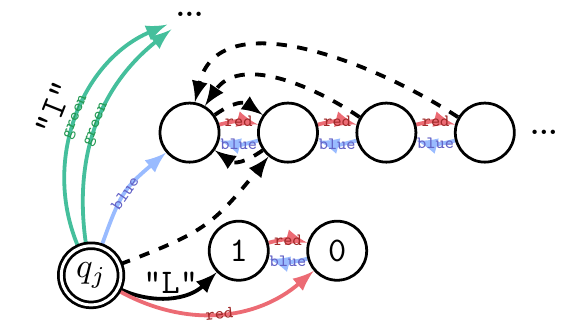}
         \caption{Symbol ``1'' written into CACHE.}
         \label{fig:ex1a}
     \end{subfigure}
     \hfill
     \begin{subfigure}[b]{0.32\textwidth}
         \centering
         \includegraphics[width=\textwidth]{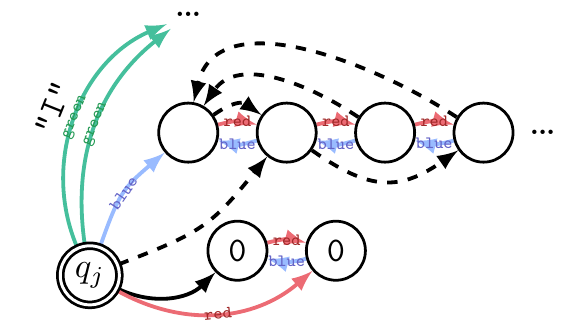}
         \caption{CACHE loaded into second block.}
         \label{fig:ex1b}
     \end{subfigure}
     \hfill
     \begin{subfigure}[b]{0.32\textwidth}
         \centering
         \includegraphics[width=\textwidth]{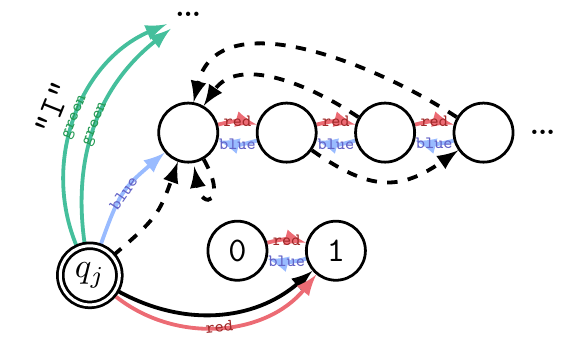}
         \caption{First block loaded into CACHE.}
         \label{fig:ex1c}
     \end{subfigure}
        \caption{Changing blocks}
        \label{fig:ex1}
\end{figure}

Figure \ref{fig:ex1} shows the process of changing the active block. In Subfigure \ref{fig:ex1a}, a rule from \ttt{Transitions} has just been applied, labelling the left node of CACHE \ttt{1}, and the unmarked edge ``\ttt{L}''. The working tape head is yet to be moved. Next, in \ttt{Simulate}, \ttt{Left} matches and \ttt{PrevBlock} is called. The procedure \ttt{Encode} loads the content of CACHE into a dashed edge within BLOCKSET. While CACHE is decremented as a ternary number, the outgoing dashed edge of the current (second) BLOCKSET node is shifted to the right. This produces the graph from Subfigure \ref{fig:ex1b}. CACHE does not represent the content of the second block anymore, but the dashed edge from the second node in BLOCKSET to the fourth does. Next we move one block to the left with \ttt{Prev}, and reposition the tape head in CACHE with \ttt{HeadRight!}. We then load the new block into CACHE. While moving the outgoing dashed edge of the current (first) BLOCKSET node to the left, the ternary number in CACHE is incremented. This results in the graph from Subfigure \ref{fig:ex1c}. We have now loaded the content of the first block into CACHE, so there is no need to store it with a dashed edge anymore. Hence the dahed edge goes from the first node in BLOCKSET to itself by convention.

% \end{example}

\subsection{Correctness}\label{subsec:correctness}

We define the \emph{configuration} of a Turing machine to consist of the input and working tapes, the position of the tape heads on those tapes, and the current state.

Let us define what graphs programs in \ttt{Sim} operate on. We call graphs as in Figure \ref{fig:graph-representation} \emph{configuration graphs}. These graphs can vary from the depicted graph in the following ways. The targets of the unlabelled green edge, dashed edges, and the unmarked edge can be any node in INPUT, BLOCKSET, and CACHE respectively. 
An exception to that is the outgoing dashed edge of the active node in BLOCKSET (target of dashed edge from the root), which is targeted towards the leftmost node in BLOCKSET since the content of that block is currently stored in CACHE. This is a convention that allows the graph representation of a configuration to be unique.
The labels of nodes in INPUT, CACHE, and the root node can be any element from the input alphabet, tape alphabet, or set of states respectively, encoded as SGP\,2 labels. 

Note that the graphs' capacity to represent tape squares is limited by the sizes of BLOCKSET and CACHE. Hence for a non-negative integer $k$, we define a function $\mathrm{enc}_k$ that encodes the configuration $S$ of a Turing machine as the configuration graph $\mathrm{enc}_k(S)$ with $3^{k+2}$ nodes in BLOCKSET and $k+2$ in CACHE. The integer $k$ corresponds to the number of times the tape is extended by the procedure \ttt{Restart}. For a given value of $k$ that is large enough, $\mathrm{enc}_k(S)$ exists and is unique.% Furthermore, $\mathrm{enc}_k$ establishes a one-to-one correspondence between configuration graphs related to $k$ and Turing machine configurations with at most $k$ non-blank working tape squares.

For a Turing machine $M$ we denote a single transition from configuration $S$ to $S'$ by $S \Rightarrow_M S'$, and the transitive reflexive closure of that relation using $\Rightarrow_M^*$. Similarly, for an SGP\,2 rule \ttt{r} we use $S \Rightarrow_{\ttt{r}} S'$ to denote a change of configuration, and $S \Rightarrow_{\ttt{P}}^* S'$ for a program \ttt{P}.

During the execution of $\ttt{Sim}(M)$, graphs that are not configuration graphs are generated. They only differ in terms of edge labels, node marks, roots, and dashed edge targets. However, these variations are temporary. Once the procedure \ttt{Simulate} has successfully terminated, the host graph is once again a configuration graph.

We say an SGP\,2 program \ttt{P!} (i.e. a loop where procedure \ttt{P} is applied as long as possible) \emph{simulates} a Turing machine $M$ of initial configuration $I$ if, for each two configurations $S$ and $S'$ of $M$ such that $I \Rightarrow_M^* S \Rightarrow_M S'$, we have $\mathrm{enc}_0(I) \Rightarrow^*_{\ttt{P!}} \mathrm{enc}_k(S) \Rightarrow^*_{\ttt{P}} \mathrm{enc}_k(S')$ for some integer $k$.

The following theorem shows that this is indeed the case for the class of programs presented in this paper. %The proof can be found in Appendix \ref{app:correctness}.% However, finding $k$ is discussed in the proof of Theorem \ref{thm:space-complexity}.

\begin{theorem}[Correctness]\label{thm:correctness}
Let $M$ be a Turing machine and \ttt{Sim}$(M)$ the corresponding SGP\,2 program. Then the subprocedure \ttt{Simulate!} simulates $M$.
\end{theorem}
\begin{proof}
Let $S$ and $S'$ be configurations of $M$ such that $I \Rightarrow_M^* S \Rightarrow_M S'$. Then some transition $\delta(q,a,x) = (p,y,X,Y)$ happened, where $p,q\in Q$, $a \in \Sigma$, $x,y \in \Gamma$, and $X,Y \in \{\mathrm{L},\mathrm{S},\mathrm{R}\}$. So the only changes from $S$ to $S'$ are that $q$ and $x$ have been updated to $p$ and $y$, and that the input and working tape heads have been shifted into direction $X$ and $Y$ respectively.

We proceed by induction on the length of $I \Rightarrow_M^* S$. For the base case we only need to show $\mathrm{enc}_k(S) \Rightarrow^*_{\ttt{P}} \mathrm{enc}_k(S')$ for some integer $k$. This also happens to be what we need to show for the induction step since $\mathrm{enc}_0(I) \Rightarrow^*_{\ttt{P!}} \mathrm{enc}_k(S)$ is given by the induction hypothesis, and the existence of a large enough $k$ by part of the proof of Theorem \ref{thm:space-complexity}.

Now it remains to show that $\mathrm{enc}_k(S) \Rightarrow^*_{\ttt{P}} \mathrm{enc}_k(S')$. The procedure \ttt{Reset} can be ignored because it is never called since we already have a large enough $k$.

When running \ttt{Simulate} on the graph $\mathrm{enc}_k(S)$, the first call is the rule set \ttt{Transitions}. A rule in that set that is guaranteed to be applicable is the one corresponding to the aforementioned Turing machine transition. It correctly updates the $q$ and $x$ to be $p$ and $y$ to match $\mathrm{enc}_k(S')$. It also moves the input tape head to the correct position. For the rest of this proof, we will show that the working tape head moves to the correct position.

If $Y=\mathrm{S}$, the working tape head is not moved, corresponding to $\mathrm{enc}_k(S)'$. Due to lack of an unmarked edge label, none of the \ttt{try} conditions in \ttt{Simulate} are applicable, and we terminate with $\mathrm{enc}_k(S')$.

The cases for $Y=\mathrm{L}$ and $Y=\mathrm{R}$ are analogous, so we will only argue for the former. When attempting to apply \ttt{MoveLeft}, if the target of the unmarked edge is not the leftmost node of CACHE, the rule set succeeds, the unmarked edge is in the correct position, and no other conditions of \ttt{try} statements in \ttt{Simulate} are applicable. Hence we terminate with $\mathrm{enc}_k(S')$.

If target of the unmarked edge is the leftmost node of CACHE however, the only \ttt{try} condition that can match is \ttt{Left}. This removes the edge label and calls \ttt{PrevBlock}. So it remains to argue that \ttt{PrevBlock} turns the host graph into $\mathrm{enc}_k(S')$.

The commands \ttt{Prev; HeadRight!} correctly position the working tape head on the rightmost node of the previous block. It remains to show that \ttt{Encode} and \ttt{Decode} preserve the working tape content according to the encoding $\mathrm{enc}_k$.

The procedure \ttt{Encode} contains a loop that, whenever it decrements the content of CACHE, it increments the content of the active block (represented by an outgoing dashed edge). This happens until the CACHE content reaches $0$, meaning the content of CACHE is correctly store in BLOCKSET. After moving to the new block, the procedure \ttt{Decode} is called. It contains a loop that, whenever it decrements the content of the active block, increments the content of CACHE. This terminates once the content of the active block has reached $0$, meaning the content of the active block has been correctly stored in CACHE. Hence the host graph is now $\mathrm{enc}_k(S')$.
\end{proof}

\section{From the Simulation to Computation Models}\label{sec:models}

Based on the \ttt{Sim} programs, we can generalise to computation models that are subsets of GP\,2, for which we can define a time measure. One obstacle is time-intensive rule matching. A second one is that when \emph{critical subprograms} (bodies of loops and conditions of \ttt{if} and \ttt{try}) fail, they need to be undone, which can be handled in different ways. The current implementation of GP\,2 reverses the relevant rule applications. The formal semantics uses a stack of graphs to keep track of the state of the host graph before entering a critical subprogram. Reversion can then be done by a simple pop operation. The downside to this is that the host graph has to be duplicated every time a loop or a branching statement is entered, which can add a polynomial factor to the time complexity of the overall program. It is also space-intensive, but only by a polynomial factor since only finitely many loop and branching statements can be nested in a command sequence. To avoid undoing, one wants failed critical subprograms to be \emph{null}, i.e. they do not change the graph state, meaning they do not need to be undone. 

Models that overcome these obstacles can be implemented to match rules in constant time, which allows us to assign unit time cost to that operation. They also make undoing redundant, allowing us to charge no time cost for that.

Another obstacle is that matching a set of rules is nondeterministic, and backtracking that nondeterminism can be costly in time or space. This can be addressed at the implementation level. The current GP\,2 compiler for instance picks the first match it finds and does not backtrack. A more general approach however is to show that within the scope of running a program on an input graph, every rule and rule set (nondeterministic call of a list of rules) can only have at most one match, effectively making the program deterministic. This property applies to the class of programs in this paper and is shown in Proposition \ref{prop:nondeterminism}.

We define an \emph{efficient model} as $\mathcal{M} = \tuple{\mathcal{P},\mathcal{I}}$, where $\mathcal{P}$ is a set of GP\,2 programs and $\mathcal{I}$ a set of GP\,2 graphs such that the following two properties are satisfied within the scope of derivation sequences starting with any $\tuple{P,I}\in \mathcal{M}$.

(1) \emph{Constant Matching}: Every rule matches in constant time.

(2) \emph{Critical Subprogram}: Every critical subprogram that fails is null.

Let us define the \emph{graph space} measure of a graph as the number of nodes plus the number of edges. We do not consider the size of labels in this paper since all labels are of constant size. This measure is uniform in that it gives unit cost to each node and edge. A discussion about this uniformity can be found at the end of Section \ref{sec:complexity}.

For an efficient model $\mathcal{M}$, we define the time complexity of a program $P$ of $\mathcal{M}$ as the maximum number of rule calls in terminating derivation sequences starting with $P$ on graphs of a given size.

\medskip
We show in this section that $\tuple{P,I}$, where $P=\{\ttt{Sim}(M) \,|\, M \textrm{ is a Turing} $ ma\-chine$\}$ and $I=\{\mathrm{CG} \,|\, \text{CG is}$ $\text{a configuration graph}\}$, is an efficient model. The constant matching property is given by Theorem \ref{thm:matching}, which holds because all rules are fast, input graphs have bounded outdegree and a bounded number of roots, and the programs preserve this boundedness. Moreover, when extending the tape with \ttt{Restart}, outgoing dashed edges of existing block nodes are removed, and one outgoing dashed edge is added for each block node, preserving bounded outdegree.
The critical subprogram property is shown by Proposition \ref{prop:null-failure}.

% \hl{There are other ways to show the constant matching property. Programs using depth-first search in} \cite{Campbell-Courtehoute-Plump22a}\hl{, like the connectedness recognition and 2-colouring programs for instance use the original version of the fast matching theorem, which restricts host graphs to bounded degree, but allows for rules where each node in the left-hand side only needs to be undirectly reachable from a root. Some rules do not even have roots, and their constant matching is given by the fact that there are only constantly many invalid matches. Programs based on reduction from the same paper, such as the tree recognition program, are proven to be efficient without restrictions on degree even.}

\medskip
The remainder of this section consists of aforementioned propositions and a lemma needed to prove one of them.

\begin{lemma}\label{lem:overhead}
    Let \ttt{P} be a critical subprogram in a graph state $S$. Assume either \ttt{P} cannot fail from state $S$, or \ttt{P} can only fail from state $S$ due its first component being a rule or rule set call that fails to match. Then if \ttt{P} fails from state $S$, \ttt{P} is null.
\end{lemma}
\begin{proof}
    If \ttt{P} cannot fail from state $S$, the lemma is trivially satisfied. Now assume \ttt{P} fails from $S$ due to its first component being a rule or rule set failing to match. Since it failed to match, the first component cannot have changed the host graph, and since it is the first, no other component can have changed the host graph.
\end{proof}

\begin{proposition}[Critical Subprogram Property]\label{prop:null-failure}
    In \ttt{Sim}$(M)$, given configuration graphs as inputs, every critical subprogram that fails is null.
\end{proposition}
\begin{proof}
    We will argue for each critical subprogram of \ttt{Sim} that Lemma \ref{lem:overhead} applies.

    The conditions of all \ttt{if} and \ttt{try} statements are either rules or rule set calls, satisfying Lemma \ref{lem:overhead}.

    Let us now argue for the loop bodies. The procedures \ttt{Erase}, \ttt{RewindCache}, \ttt{HeadLeft} as well as \ttt{HeadRight} are rule sets, and \ttt{rewind\_blockset} and \ttt{direct} are rules, satisfying Lemma \ref{lem:overhead}.
    
    \ttt{Reset}: The rule \ttt{overflow} can fail, but the other component cannot since it is a \ttt{try} statement whose branches only contain \ttt{break} which cannot fail.
    
    \ttt{Increment}: This consists of a \ttt{try} statement, so only the branches can fail. \ttt{Finish} cannot fail because \ttt{CacheInit} is always called before this loop, providing a match. The rule \ttt{overflow} cannot fail because it is only called when \ttt{Inc} fails and the labels of nodes in CACHE are either \ttt{0}, \ttt{1}, or \ttt{2}. The \ttt{try} statement cannot fail because its branches cannot fail.
    
    \ttt{Decrement}: The reasoning is analogous to that in the previous paragraph.
    
    \ttt{Decoding}: The \ttt{try} statement cannot fail because its branches cannot fail. For the remainder of this paragraph, we argue that \ttt{CacheInc} cannot fail either. \ttt{Increment!} cannot fail because it is a loop. \ttt{CacheInit} always succeeds because the target of the red edge originating from the unmarked root is unmarked because of the structure of the input graph, and because previous calls of \ttt{CacheInc} and \ttt{CacheDec} turn the only marked rooted node in this part of the graph unmarked and unrooted with \ttt{Finish}.
    
    \ttt{Encoding}: The reasoning is analogous to that in the previous paragraph.
    
    The body of the loop \ttt{(copy; try Undirect)!} contained in \ttt{ResetBlockset}: The rule \ttt{copy} is allowed to fail, and \ttt{try Undirect} cannot fail since it is a \ttt{try} statement.
    
    \ttt{Simulate}: \ttt{Transition} is a rule set at the start of the body and hence allowed to fail. The rest consists of \ttt{try} statements whose branches call \ttt{PrevBlock} and \ttt{NextBlock}. \ttt{Prev} and \ttt{Next} cannot fail because we assume that BLOCKSET in the input graph is large enough to accommodate the execution of the Turing machine. \ttt{Encode} and \ttt{Decode} do not fail, which we will argue for the rest of this paragraph. \ttt{EncodeInit} and \ttt{DecodeInit} find a match because of the structure of the input graph and because previous calls of \ttt{Encode} and \ttt{Decode} leave BLOCKSET unmarked and unrooted. \ttt{Encoding!} and \ttt{Decoding!} are loops and hence cannot fail. \ttt{Update} has a match regardless of the blue root's location within BLOCKSET.
\end{proof}

\begin{proposition}[Unique Matches]\label{prop:nondeterminism}
    In \ttt{Sim}$(M)$, given configuration graphs as inputs, whenever a rule or rule set is called, there is at most one match.
\end{proposition}
\begin{proof}
    In each rule set, if the left-hand side of two rules have the same structure, their labels are different. In \ttt{Transitions} in particular, no two rules have the same left-hand side since we consider deterministic Turing machines. Among rules with different labels but the same structure, at most one can match, namely the one that has the same labels as the host graph. Among rules with different structures and the right labels, at most one can match because they differ by an edge or node that is unique in host graphs. Furthermore, it is easy to check that each rule can only have at most one match in the host graph. Each rule contains a root with a unique combination of mark and label. And for each node, outgoing edges can be distinguished in the same way.
\end{proof}

\section{Results on Time and Space Complexity}
\label{sec:complexity}

In this section, we present theorems on the time and space complexities of the simulation.

\begin{theorem}[Time Complexity]\label{thm:time-complexity}
Every Turing machine $M$ of time complexity $t(n)$ is simulated by \ttt{Sim}$(M)$ in time $\mathrm{O}(t^2(n))$, where $n$ is the size of the input.
\end{theorem}
\begin{proof}
% First of all, Proposition \ref{prop:null-failure} guarantees that there is no linear time overhead when entering loops, \ttt{if} statements, or \ttt{try} statements.

% Let us argue that all rules match in constant time by applying Theorem \ref{thm:matching}. All rules are fast since nodes in the left-hand side are reachable from a root, and since no rule has variables or conditions. The host graphs have a constant number of roots since the input graphs have a single one, and whenever one is introduced by \ttt{CacheInc}, \ttt{CacheDec}, \ttt{Encode}, or \ttt{Decode}, it is also removed by the same procedure. The outdegree of the input graph is bounded. None of the rules modify the source of an edge, so outdegree is preserved.

Given the discussion in Section \ref{sec:models}, we can assign unit time to rule and rule set calls and argue time complexity to be the number of such calls.

First, we will show that simulating one step of $M$ (not counting restarting the simulation) takes $\mathrm{sim}(n) = \mathrm{O}(s(n))$ time. We consider the size of CACHE from the final simulation since it provides an upper bound. The only sources of non-constant time are \ttt{PrevBlock} and \ttt{NextBlock}. Their complexity is the worst of the loops \ttt{Encoding!}, \ttt{Decoding!}, \ttt{HeadLeft!}, and \ttt{HeadRight!}. The latter two simply traverse CACHE, which takes $\mathrm{O}(s(n))$ time. The former two decrement/increment CACHE as a ternary counter. Their time complexity is proportional to the number of digit operations it takes to decrement the counter from the number $s(n)$ all the way to $0$. The rightmost digit is modified $s(n)$ times, the next one $\frac{1}{3}s(n)$ times, the one after $\frac{1}{3} \cdot \frac{1}{3} s(n)$ times, and so on. So the total number of digit operations is $\sum_{k=0}^{\log s(n)} s(n)\left( \frac{1}{3} \right)^k $. Using properties of the geometric series, one can see that this is $\mathrm{O}(s(n))$.

In this paragraph, we will show that resetting the simulation and extending the tape takes $r(n) = \mathrm{O}(s(n))$ time. Consider the loops of \ttt{Restart}. Both \ttt{RewindCache!} and \ttt{Erase!} traverse CACHE and hence take time $\mathrm{O}(\log s(n))$. The loop \ttt{(copy; try Undirect)!} in \ttt{ResetBlockset} traverses BLOCKSET and thus takes $\mathrm{O}(s(n))$ time. And \ttt{direct!} traverses the extended BLOCKSET in $\mathrm{O}(3\cdot s(n))=\mathrm{O}(s(n))$ time.

Next, we will show that the number of times the simulation is restarted is $l(n)=\mathrm{O}( \log s(n))$. The final size of the tape of $M$ is $\mathrm{O}(s(n) \log s(n))$. Since the number of represented squares is tripled in each step, the number of steps is $\mathrm{O}(\log ( s(n) \log s(n) ))$. Using the formula for the logarithm of a product, one can simplify this to $\mathrm{O}( \log s(n))$.

The total time complexity can be bounded by $\textrm{reset}(n) + \textrm{simulation}(n)$, where $\textrm{reset}(n) = l(n) \cdot r(n)$ is the total cost of all resets, and $\textrm{simulation}(n) = l(n) \cdot t(n) \cdot \mathrm{sim}(n)$ the total cost of simulating $M$ across all resets. Using results from previous paragraphs, we get $\textrm{reset}(n) = \mathrm{O}(s(n) \log s(n))$ and $\textrm{simulation}(n) = \mathrm{O}(t(n) \cdot s(n) \log s(n))$. Since space complexity $s(n) \log s(n)$ can be bounded by time complexity $t(n)$, the entire simulation takes $\mathrm{O}(t^2(n))$ time.

\end{proof}

\begin{theorem}[Space Complexity]\label{thm:space-complexity}
Every Turing machine $M$ of space complexity $\mathrm{O}(s(n) \log s(n))$ is simulated by \ttt{Sim}$(M)$ in graph space $O\left( s(n) \right)$, where $n$ is the size of the input.
\end{theorem}
\begin{proof}
During the execution of \ttt{Sim}$(M)$, nodes and edges are only created by \ttt{setup} and \ttt{Restart}, and none are ever deleted. The numbers of nodes and edges only differ by a constant factor since the outdegree is bounded, so we will argue for space complexity using number of nodes only. 

Initially, after application of \ttt{setup}, BLOCKSET has $b(n) = 3^2$ nodes, and CACHE $c(n)=2$. Then, each application of \ttt{Restart} adds one to $c(n)$ and triples $b(n)$. So after $k$ iterations, we have $b(n) = 3^{2+k}$ and $c(n)=2 + k$.

The Turing machine needs $S(n) = \mathrm{O}(s(n) \log s(n))$ tape squares. This means that there are positive integers $n_0$ and $c$ such that $S(n) \leq c\, s(n) \log s(n)$ for all $n \geq n_0$. So for all $n$ we can say $S(n) \leq c\, s(n) \log s(n) + m$, where $m=\max_{n\in\{0,\dots,n_0\}}S(n)$, a constant.

Assume \ttt{Restart} is called $k= \log s(n) -2 +d$ times, where $d=\max(m,\,\log c)$. 
For that value of $k$, we have $c(n) b(n)=3^d s(n) \log s(n) + 3^d s(n) \,d$. By definition we have $d\geq m$ and $3^d \geq c$. Furthermore, we have $3^d s(n) \geq 1$. Hence we get $c(n) \cdot b(n) \geq c\, s(n) \log s(n) +m \geq S(n)$.
So for this value of $k$, the graph can store enough tape squares to execute $M$.

With the aforementioned value of $k$, the number of nodes in this graph is $c(n)+b(n) = 3^d s(n) + \log s(n) + d$, which is  $\mathrm{O}(\log s(n) + s(n))=\mathrm{O}(s(n))$.

Hence the number of nodes of the graph that is created is bounded by $c(n)+b(n)$, and hence in $\mathrm{O}(s(n))$.
\end{proof}

One might wonder why this space compression is not possible on random access machines. GP\,2 does have a C implementation after all. The reason for this is related to how graphs are represented in random access machines (RAMs), and how much space that takes. Graph edges are usually implemented as pointers. However the size of pointer addresses grows logarithmically with the number of nodes, since these addresses are usually stored as binary numbers. So in the context of RAMs, it does not seem very accurate to assign unit cost to edges. To take this into account, one can use \emph{logarithmic space}, in which graphs of $s(n)$ nodes are assigned a cost of $s(n) \log s(n)$.

If we use logarithmic space on our model, space compression is nullified since the simulation then has the same asymptotic space complexity as the machine it simulates. This puts into question whether uniform or logarithmic space should be used, which is discussed by van Emde Boas \cite{VanEmdeBoas89a}. One may want to charge more than unit space since in RAMs, edges are represented by pointers whose size grows with the number of nodes. A related issue can be found in the time measure of RAM models when programs have to deal with large integers.

\section{Conclusion}

In GP\,2, we have found the same space compression phenomenon that SMMs and KUMs exhibit. Unlike the other computation models, GP\,2 uses rule-based graph transformation. The compression happens due to use of a graph-based data structure to encode the tape of a Turing machine, which is more space-efficient with respect to uniform measures.

Using this simulation, we have identified efficient computation models that are subsets of GP\,2. Requiring these models to have certain properties allows us to define rigorous complexity measures.

In future work, the simulation could be refined in the way Luginbuhl \cite{Luginbuhl-Loui93a} refines the one by van Emde Boas \cite{VanEmdeBoas89a}, namely with a more complex data structure that allows for real-time simulation with some form of space compression.

Furthermore, the simulation could be extended to nondeterministic Turing machines. Extending the simulation would be straightforward, but the complexity results would need further proofs.

One could also investigate what measurable impact this space compression has on a practical example. Graph programs working on linked lists are the most likely to benefit from this paper's space compression. As a case study, one could pick a list sorting algorithm, implement it on integer lists both with and without space compression, and compare their memory usage.

% TODO: conceivable applications of this result: presumably graph programs working on linked lists, for instance sorting programs on integer lists, measuring the memory usage of such a list program and compare it to a program that uses the compression technique of the paper

% One could attempt to extend the simulation to a larger efficient model that is complete in the sense of being able to compute graph functions. Our simulation allows for Turing-completeness, which is weaker for a graph programming model.

\nocite{*}
\bibliographystyle{eptcs}
\bibliography{generic}

\appendix

\end{document}

%% file: macros.tex
\usepackage{amsmath,amssymb,amsthm}

\newtheorem{theorem}{Theorem}
\newtheorem{lemma}{Lemma}
\newtheorem{proposition}{Proposition}

\newtheorem{definition}{Definition}
\usepackage{stmaryrd}

\newcommand{\dder}{\Rightarrow}

\newcommand{\mtt}{\mathtt}
\newcommand{\mrm}{\mathrm}

\newcommand{\failrm}{\mathrm{fail}}

\renewcommand{\L}{\mathcal{L}}
\newcommand{\tuple}[1]{\langle#1\rangle}

\newcommand{\ifte}[3]{\mathtt{if}\ #1\ \mathtt{then}\ #2\ \mathtt{else}\ #3}

\newcommand{\tryte}[3]{\mathtt{try}\ #1\ \mathtt{then}\ #2\ \mathtt{else}\ #3}

\newcommand\sdots{\hbox to 1em{.\hss.\hss.}}

\newcommand{\ttt}{\texttt}
\newenvironment{tttt}{\ttfamily}{\par}

\usepackage{tabularx,environ}
\makeatletter
\newcommand{\specinput}[1]{\gdef\@specinput{#1}}%
\newcommand{\specoutput}[1]{\gdef\@specoutput{#1}}
\NewEnviron{spec}{
  \BODY
  \par\addvspace{.25\baselineskip}
  \noindent
  \begin{tabularx}{\textwidth}{@{\hspace{\parindent}} l X c}
    \textbf{Input:} & \@specinput \\
    \textbf{Output:} & \@specoutput
  \end{tabularx}
  \par\addvspace{.25\baselineskip}
}
\makeatother

\newcommand{\G}{\mathcal{G}}

\usepackage{cases}

\usepackage{subcaption}

\usepackage{breqn}

\usepackage[linesnumbered,ruled,vlined]{algorithm2e}

\usepackage{tikz}
\usetikzlibrary{cd}
\usetikzlibrary{shapes.misc}
\tikzset{>=latex}
\tikzset{every edge/.style={draw,thick}}
\usepackage{adjustbox}

\usepackage{xcolor}
\definecolor{bc-green-light}{RGB}{69, 191, 156}
\definecolor{bc-green-dark}{RGB}{37, 154, 85}
\definecolor{bc-blue-light}{RGB}{153, 187, 255}
\definecolor{bc-blue-dark}{RGB}{100, 99, 198}
\definecolor{bc-red-light}{RGB}{236, 107, 116}
\definecolor{bc-red-dark}{RGB}{161, 41, 41}
\definecolor{bc-pink-light}{RGB}{239, 161, 193}
\definecolor{bc-pink-dark}{RGB}{192, 62, 156}
\definecolor{bc-grey-light}{RGB}{196, 192, 200}
\definecolor{bc-grey-dark}{RGB}{135, 135, 135}

\tikzset{gp2 node/.style={draw, rounded rectangle, minimum height=.55cm, minimum size=6mm, thick}}
\tikzset{gp2 red node/.style={draw, rounded rectangle, minimum height=.55cm, minimum size=6mm, thick, fill=bc-red-light, TeXture={\tiny{\textcolor{bc-red-dark}{red}}}, TeXture sep=2pt}}
\tikzset{gp2 blue node/.style={draw, rounded rectangle, minimum height=.55cm, minimum size=6mm, thick, fill=bc-blue-light, TeXture={\tiny{\textcolor{bc-blue-dark}{blue}}}, TeXture sep=2pt}}
\tikzset{gp2 green node/.style={draw, rounded rectangle, minimum height=.55cm, minimum size=6mm, thick, fill=bc-green-light, fill=bc-green-light, TeXture={\tiny{\textcolor{bc-green-dark}{green}}}, TeXture sep=2pt}}
\tikzset{gp2 grey node/.style={draw, rounded rectangle, minimum height=.55cm, minimum size=6mm, thick, fill=bc-grey-light, TeXture={\tiny{\textcolor{bc-grey-dark}{grey}}}, TeXture sep=2pt}}
\tikzset{gp2 any node/.style={draw, rounded rectangle, minimum height=.55cm, minimum size=6mm, thick, fill=bc-pink-light, TeXture={\tiny{\textcolor{bc-pink-dark}{any}}}, TeXture sep=2pt}}

\tikzset{gp2 root/.style={draw, rounded rectangle, minimum height=.55cm, minimum size=6mm, thick, double, double distance=0.3mm}}
\tikzset{gp2 red root/.style={draw, rounded rectangle, minimum height=.55cm, minimum size=6mm, thick, double, double distance=0.3mm, fill=bc-red-light, TeXture={\tiny{\textcolor{bc-red-dark}{red}}}, TeXture sep=2pt}}
\tikzset{gp2 blue root/.style={draw, rounded rectangle, minimum height=.55cm, minimum size=6mm, thick, double, double distance=0.3mm, fill=bc-blue-light, TeXture={\tiny{\textcolor{bc-blue-dark}{blue}}}, TeXture sep=2pt}}
\tikzset{gp2 green root/.style={draw, rounded rectangle, minimum height=.55cm, minimum size=6mm, thick, double, double distance=0.3mm, fill=bc-green-light, TeXture={\tiny{\textcolor{bc-green-dark}{green}}}, TeXture sep=2pt}}
\tikzset{gp2 grey root/.style={draw, rounded rectangle, minimum height=.55cm, minimum size=6mm, thick, double, double distance=0.3mm, fill=bc-grey-light, TeXture={\tiny{\textcolor{bc-grey-dark}{grey}}}, TeXture sep=2pt}}
\tikzset{gp2 any root/.style={draw, rounded rectangle, minimum height=.55cm, minimum size=6mm, thick, double, double distance=0.3mm, fill=bc-pink-light, TeXture={\tiny{\textcolor{bc-pink-dark}{any}}}, TeXture sep=2pt}}

\tikzset{every edge quotes/.style={font=\tiny}}
\tikzset{gp2 edge/.style={->,thick,line width=1.1pt}}
\tikzset{gp2 red edge/.style={->,bc-red-light,"red"{bc-red-dark},sloped,line width=1.1pt}}
\tikzset{gp2 blue edge/.style={->,bc-blue-light,"blue"{bc-blue-dark},sloped,line width=1.1pt}}
\tikzset{gp2 green edge/.style={->,bc-green-light,"green"{bc-green-dark},sloped,line width=1.1pt}}
\tikzset{gp2 dashed edge/.style={->,thick,dashed,line width=1.1pt}}
\tikzset{gp2 any edge/.style={->,bc-pink-light,"any"{bc-pink-dark},sloped,line width=1.1pt}}

\tikzset{gp2 bi edge/.style={-,thick,line width=1.1pt}}
\tikzset{gp2 red bi edge/.style={-,line width=5pt,bc-red-light,"red"{bc-red-dark},sloped,line width=1.1pt}}
\tikzset{gp2 blue bi edge/.style={-,line width=5pt,bc-blue-light,"blue"{bc-blue-dark},sloped,line width=1.1pt}}
\tikzset{gp2 green bi edge/.style={-,line width=5pt,bc-green-light,"green"{bc-green-dark},sloped,line width=1.1pt}}
\tikzset{gp2 dashed bi edge/.style={-,thick,dashed,line width=1.1pt}}
\tikzset{gp2 any bi edge/.style={-,line width=5pt,bc-pink-light,"any"{bc-pink-dark},sloped,line width=1.1pt}}

\tikzset{none/.style={}}

\usepackage{pgfplots}
\pgfdeclarelayer{edgelayer}
\pgfdeclarelayer{nodelayer}
\pgfsetlayers{edgelayer,nodelayer,main}
\pgfplotsset{compat=1.16}

\newcommand*\tikzTeXture[1]{% %% <- Does the actual filling
  \begingroup
    \setbox0=\vbox spread \pgfkeysvalueof{/tikz/TeXture y sep}{\vfil%
               \hbox spread \pgfkeysvalueof{/tikz/TeXture x sep}{\hfil#1\hfil}\vfil}%
    \def\fillareasize{\pgfpointdiff %% <- size of area to be filled
      {\pgfpointanchor{path picture bounding box}{south west}}%
      {\pgfpointanchor{path picture bounding box}{north east}}}%
    \pgfextractx{\dimen0}{\fillareasize}% %% <- width of area to be filled
    \pgfextracty{\dimen2}{\fillareasize}% %% <- height of area to be filled
    \leavevmode\hbox to \dimexpr\dimen0+\wd0{%
      \cleaders\vbox to \dimexpr\dimen2+\ht0+\dp0{
        \cleaders\box0\vfil
      }\hfil
    }%
  \endgroup
}
\tikzset{TeXture/.style={path picture={
  \node[anchor=center,text width=,text height=]
        at (path picture bounding box.center) {\tikzTeXture{#1}};}
}}
\tikzset{TeXture/.default=\TeX}
\tikzset{TeXture x sep/.initial=0pt}
\tikzset{TeXture y sep/.initial=0pt}
\tikzset{TeXture sep/.style={/tikz/TeXture x sep=#1,/tikz/TeXture y sep=#1}}

%% file: Figures/gp2/label-grammar.tex
\begin{figure}[h]
\centering
\begin{tabular}{lcl}
Label & ::= & Atom [Mark] \\
Atom & ::= & Integer $\mid$ \ttt{"L"} $\mid$ \ttt{"R"} $\mid$ \ttt{"I"} $\mid$ \ttt{empty} \\
Mark & ::= & red $\mid$ green $\mid$ blue $\mid$ grey $\mid$ dashed
\end{tabular}
\caption{Abstract syntax of host and rule graph labels \label{fig:label-grammar}}
\end{figure}

%% file: Figures/gp2/program-syntax.tex
\begin{figure}[h]
\centering
\begin{tabular}{lcl}
Prog & ::= & Decl \{Decl\} \\
Decl & ::= & RuleDecl $\mid$ ProcDecl $\mid$ MainDecl \\
ProcDecl & ::= & ProcId `=' ComSeq \\
MainDecl & ::= & \ttt{Main} `=' ComSeq \\
ComSeq & ::= & Com \{`;' Com\} \\
Com & ::= & RuleSetCall $\mid$ ProcCall \\
&& $\mid$ \ttt{if} ComSeq \ttt{then} ComSeq [\ttt{else} ComSeq] \\
&& $\mid$ \ttt{try} ComSeq [\ttt{then} ComSeq] [\ttt{else} ComSeq] \\
&& $\mid$ ComSeq `{!}' $\mid$ `(' ComSeq `)' $\mid$ \ttt{break}  \\
RuleSetCall & ::= & RuleId $\mid$ `\{' [RuleId \{`,' RuleId\}] `\}' \\
ProcCall & ::= & ProcId 
\end{tabular}
\caption{Abstract syntax of Small GP\,2 programs \label{fig:command-syntax}}
\end{figure}

%% file: Figures/gp2/core-semantics.tex
\begin{figure}[h]
\begin{center}
\scalebox{.85}{
\begin{tabular}{l}
$\mathrm{[call_1]}$ $\frac{\displaystyle G \dder_R H}{\displaystyle\tuple{R,\,G} \to H}$ 
% &
\quad
$\mathrm{[call_2]}$ $\frac{\displaystyle G \not\dder_R}{\displaystyle\tuple{R,\,G} \to \failrm}$
\\\\

$\mathrm{[seq_1]}$ $\frac{\displaystyle \tuple{P,\, G} \to \tuple{P',\, H}}{\displaystyle \tuple{P;Q,\, G} \to \tuple{P';Q,\, H}}$ 
% &
\quad
$\mathrm{[seq_2]}$ $\frac{\displaystyle \tuple{P,\, G} \to H}{\displaystyle \tuple{P;Q,\, G}\to \tuple{Q,\, H}}$
% &
\quad
$\mathrm{[seq_3]}$ $\frac{\displaystyle \tuple{P,\, G} \to \failrm}{\displaystyle \tuple{P;Q,\, G}\to \failrm}$
\\\\

$\mathrm{[if_1]}$ $\frac{\displaystyle \tuple{C,\, G} \to^+ H}{\displaystyle \tuple{\ifte{C}{P}{Q},\, G}\to \tuple{P,\, G}}$
% &
\quad
$\mathrm{[if_2]}$ $\frac{\displaystyle \tuple{C,\, G} \to^+ \failrm}{\displaystyle \tuple{\ifte{C}{P}{Q},\, G} \to \tuple{Q,\, G}}$
\\\\

$\mathrm{[try_1]}$ $\frac{\displaystyle \tuple{C,\, G} \to^+ H}{\displaystyle \tuple{\tryte{C}{P}{Q},\, G}\to \tuple{P,\, H}}\quad$
% &
\quad
$\mathrm{[try_2]}$ $\frac{\displaystyle \tuple{C,\, G} \to^+ \failrm}{\displaystyle \tuple{\tryte{C}{P}{Q},\, G} \to \tuple{Q,\, G}}$

\\\\

$\mathrm{[alap_1]}$ $\frac{\displaystyle \tuple{P,\, G} \to^+ H}{\displaystyle \tuple{P!,\, G} \to \tuple{P!,\, H}}$
% &
\quad
$\mathrm{[alap_2]}$ $\frac{\displaystyle \tuple{P,\, G} \to^+ \failrm}{\displaystyle \tuple{P!,\, G} \to G}$
% &
\quad
$\mrm{[alap_3]}$ $\frac{\displaystyle \tuple{P,\, G} \to^* \tuple{\mtt{break}, H}}{\displaystyle \tuple{P!,\, G} \to H}$
\\\\

$\mrm{[break]}$ $\tuple{\mtt{break}; P,\, G} \to \tuple{\mtt{break},\, G}$

% & & \\\\

\end{tabular}
}
\end{center}
\vspace{-0.333333em}
\caption{Inference rules for Small GP\,2 core commands}
\label{fig:core-semantics}
\end{figure}

%% file: Figures/Examples/working-tape.tex
\begin{figure}
  \centering
  \includegraphics[scale=1.5]{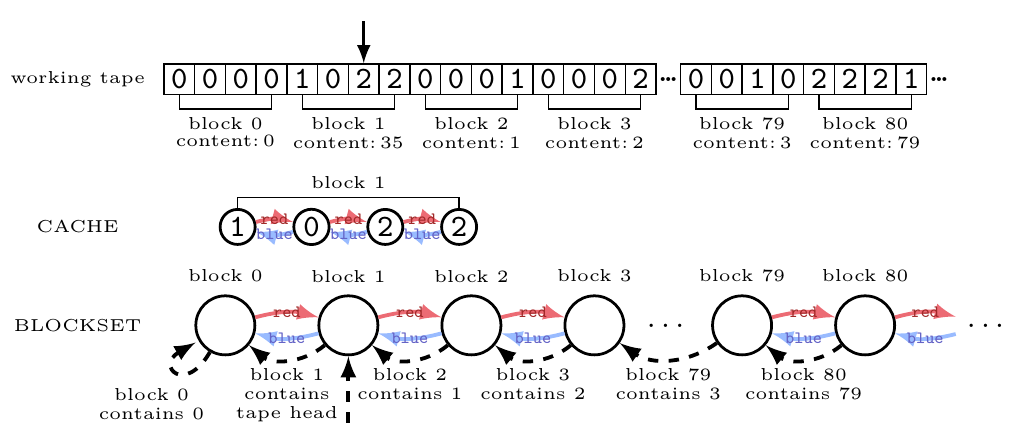}
  \caption{Graph representation of the working tape of a Turing machine.}
  \label{fig:working-tape}
\end{figure}

%% file: Figures/Examples/example.tex
\begin{figure}[h!]
  \centering
  \includegraphics[scale=1]{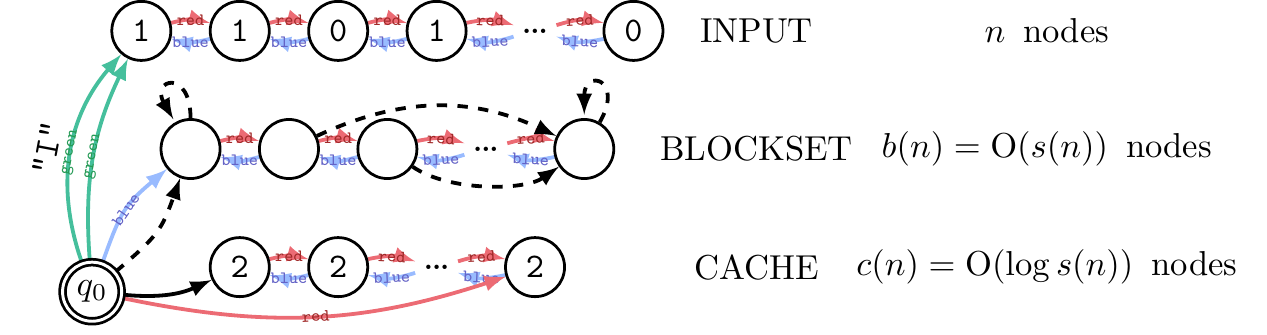}
%   \begin{subfigure}{.45\textwidth}
%     \includegraphics[scale=.85]{Figures/Examples/tm-tape/tm-tape.pdf}
%     \caption{Nonempty section of the tape}
%     \label{fig:tm-tape}
%   \end{subfigure}%
%   \begin{subfigure}{.55\textwidth}
%     \includegraphics{Figures/Examples/graph-structure/graph-structure.pdf}
%     \caption{Graph representing the tape globally in $B$ and locally in $C$}
%     \label{fig:graph-structure}
%   \end{subfigure}
  \caption{Graph representation of the initial configuration of a Turing machine of space complexity $\mathrm{O}(s(n) \log s(n))$.}
  \label{fig:graph-representation}
  \end{figure}

%% file: Figures/Simulation/simulation.tex
\begin{figure}
  \centering\scriptsize
  \fbox{\begin{minipage}{.98\textwidth} 
  \begin{tttt}
    Main = setup; Simulate!
  
    Simulate = Transitions; try MoveLeft; try MoveRight;

    \phantom{...........}try Left then PrevBlock; try Right then NextBlock; try Flag then Restart
    
  \end{tttt}
  
   \medskip
  $\ttt{MoveLeft} = \bigg\{ \,
    \raisebox{-.55\height}{\includegraphics[scale=.9]{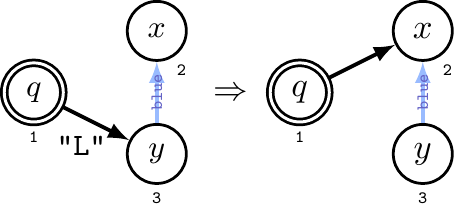}}
  \,:\,  q \in Q,\, x,y \in \Gamma\, \bigg\} $
  
  \medskip
  $\ttt{MoveRight} = \bigg\{ \,
    \raisebox{-.55\height}{\includegraphics[scale=.9]{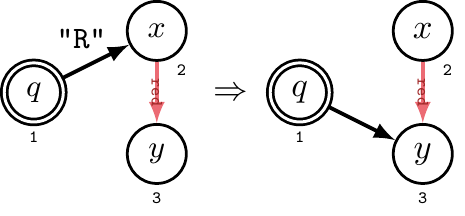}}
  \,:\,  q \in Q,\, x,y \in \Gamma\, \bigg\} $
  \quad
  $\ttt{Left} = \bigg\{ \,
  \raisebox{-.55\height}{\includegraphics[scale=.9]{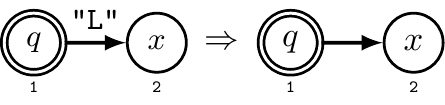}}
  \,:\, q \in Q ,\, x \in \Gamma\, \bigg\} $
  
  \medskip
  $\ttt{Right} = \bigg\{ \,
  \raisebox{-.55\height}{\includegraphics[scale=.9]{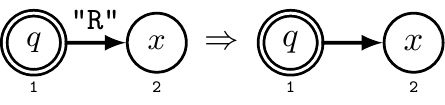}}
  \,:\, q \in Q ,\, x \in \Gamma\, \bigg\} $
  \quad$\ttt{Flag} = \bigg\{ \,
  \raisebox{-.55\height}{\includegraphics[scale=.9]{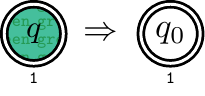}}
  \,:\, q \in Q \, \bigg\} $

  \normalsize
  
  \end{minipage}}
    \caption{SGP\,2 program $\ttt{Sim}(M)$ that simulates Turing machine $M$.}
    \label{fig:simulation}
  \end{figure}

%% file: Figures/Simulation/transitions.tex
\begin{figure}
  \centering
%   \fbox{\begin{minipage}{.98\textwidth} 
%   \begin{tttt}
%     Main = Setup; Simulate!
  
%     Simulate = try Left then PrevBlock; try Right then NextBlock; 
    
%     \phantom{...........}Transitions; try MoveLeft; try MoveRight
    
%     % MoveLeft = LTransitions; left!; PrevBlock
    
%     % MoveRight = RTransitions; right!; NextBlock

%     % \bigskip
%     % \begin{tabular}{ p{60mm} p{60mm} }
%     %   \includegraphics{Figures/Block/next-block/next-block.pdf}&
%     %   \includegraphics{Figures/Block/prev-block/prev-block.pdf}
%     % \end{tabular}
    
%   \end{tttt}
  
%   \medskip
%   $\ttt{Left} = \bigg\{ \,
%   \raisebox{-.55\height}{\includegraphics{Figures/Simulation/left/left.pdf}}
%   \,:\, q \in Q \, \bigg\} $\quad
%   $\ttt{Right} = \bigg\{ \,
%   \raisebox{-.55\height}{\includegraphics{Figures/Simulation/right/right.pdf}}
%   \,:\, q \in Q \, \bigg\} $
  
% %   \medskip
% %   $\ttt{Transitions} = \ttt{STransitions} \,\cup\, \ttt{LTransitions} \,\cup\, \ttt{RTransitions}$  
  
% %   $\ttt{STransitions} = \ttt{SSTransitions} \,\cup\, \ttt{LSTransitions} \,\cup\, \ttt{RSTransitions}$ 
  
% %   $\ttt{LTransitions} =  \ttt{SLTransitions} \,\cup\, \ttt{LLTransitions} \,\cup\, \ttt{RLTransitions} $ 
  
% %   $\ttt{RTransitions} =  \ttt{SRTransitions} \,\cup\, \ttt{LRTransitions} \,\cup\, \ttt{RRTransitions}$
  
%      \medskip
\scriptsize
  %\bigg\{\,$ %q.a.x.p.y.X.Y \,:\, \delta(q,a,x) = (p,y,X,Y),\, X,Y\in \{\textrm{L},\textrm{R},\textrm{S}\},\, b\in\Sigma \,\}$  \mathrm{LL} \cup \mathrm{LS} \cup \mathrm{LR} \cup \mathrm{SL} \cup \mathrm{SS} \cup \mathrm{SR} \cup \mathrm{RL} \cup \mathrm{RS} \cup \mathrm{RR} =
% \renewcommand\arraystretch{3}
\medskip
  \begin{tabular}{ | p{39.5mm} | p{39.5mm} | p{39.5mm} | }
  \hline 
   \multicolumn{3}{|l|}{} \\ [-.5em]
  \multicolumn{3}{|l|}{ $\ttt{Transitions} =  \bigcup_{X,Y\in\{\mathrm{L},\mathrm{S},\mathrm{R}\}}XY$} \\ [.5em]
  
    \hline
    & & \\ [-.5em]
    LL = $\bigg\{$ & LS = $\bigg\{$ & LR = $\bigg\{$\\
    \includegraphics[scale=.9]{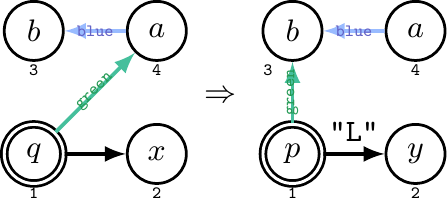}&
    \includegraphics[scale=.9]{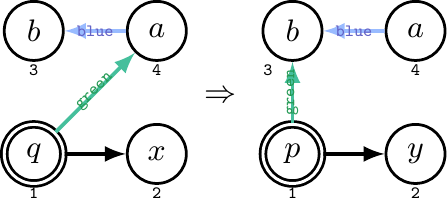}&
    \includegraphics[scale=.9]{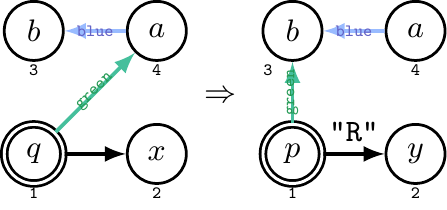}\\ 
    
    $: \delta(q,a,x) = (p,y,\mathrm{L},\mathrm{L}), b\in\Sigma \bigg\}$  & 
    $: \delta(q,a,x) = (p,y,\mathrm{L},\mathrm{S}), b\in\Sigma \bigg\}$  & 
    $: \delta(q,a,x) = (p,y,\mathrm{L},\mathrm{R}), b\in\Sigma \bigg\}$  \\ [1em] \hline
    
    & & \\ [-.5em]
    SL = $\bigg\{$ & SS = $\bigg\{$ & SR = $\bigg\{$\\
    
    \includegraphics[scale=.9]{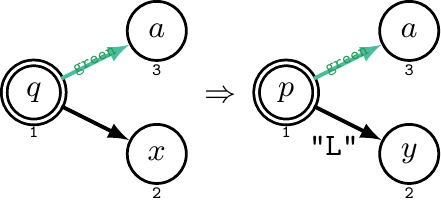}&
    \includegraphics[scale=.9]{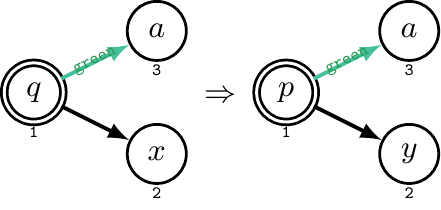}&
    \includegraphics[scale=.9]{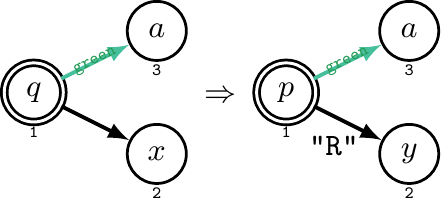}\\ 
    
    $: \delta(q,a,x) = (p,y,\mathrm{S},\mathrm{L}), b\in\Sigma \bigg\}$  & 
    $: \delta(q,a,x) = (p,y,\mathrm{S},\mathrm{S}), b\in\Sigma \bigg\}$  & 
    $: \delta(q,a,x) = (p,y,\mathrm{S},\mathrm{R}), b\in\Sigma \bigg\}$  \\ [1em] \hline
    
    & & \\ [-.5em]
    RL = $\bigg\{$ & RS = $\bigg\{$ & RR = $\bigg\{$\\
    
    \includegraphics[scale=.9]{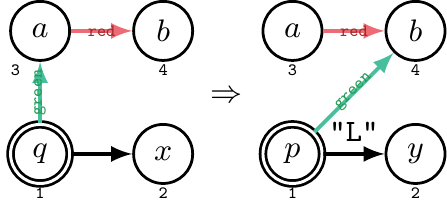}&
    \includegraphics[scale=.9]{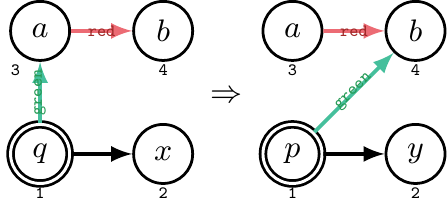}&
    \includegraphics[scale=.9]{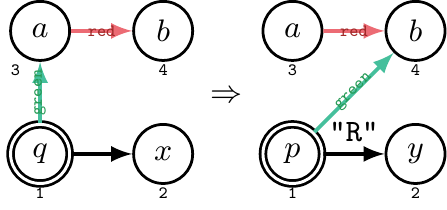}\\ 
    
    $: \delta(q,a,x) = (p,y,\mathrm{R},\mathrm{L}), b\in\Sigma \bigg\}$  & 
    $: \delta(q,a,x) = (p,y,\mathrm{R},\mathrm{S}), b\in\Sigma \bigg\}$  & 
    $: \delta(q,a,x) = (p,y,\mathrm{R},\mathrm{R}), b\in\Sigma \bigg\}$  \\ [1em] \hline
  \end{tabular}

  \normalsize
%   \end{minipage}}
    \caption{Rule set \ttt{Transitions} that models the transitions of a Turing Machine.}
    \label{fig:transitions}
  \end{figure}

%% file: Figures/Block/block.tex
\begin{figure}
  \centering\scriptsize
  \fbox{\begin{minipage}{.98\textwidth} 
  \begin{tttt}
    NextBlock = Encode; try Next then (HeadLeft!; Decode) else (SetFlag; break)
    
    % \phantom{...}
    
    PrevBlock = Encode; Prev; HeadRight!; Decode

    \medskip
    $\ttt{Next} = \bigg\{ 
  \raisebox{-.55\height}{\includegraphics[scale=.9]{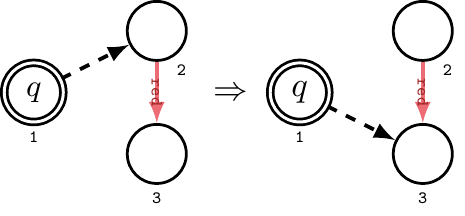}}
  : q \in Q \bigg\} \;$
  \quad
  $\ttt{Prev} = \bigg\{ 
  \raisebox{-.55\height}{\includegraphics[scale=.9]{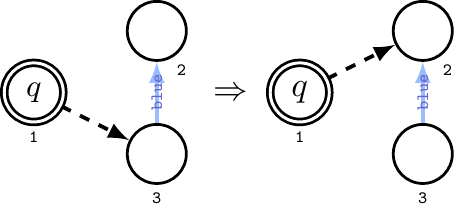}}
  : q \in Q \bigg\} $
  
  $\ttt{SetFlag} = \bigg\{ \,
    \raisebox{-.55\height}{\includegraphics[scale=.9]{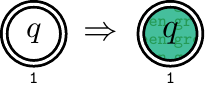}}
  \,:\,  q \in Q\, \bigg\} $
 \quad 
  $\ttt{HeadLeft} = \bigg\{ \,
    \raisebox{-.55\height}{\includegraphics[scale=.9]{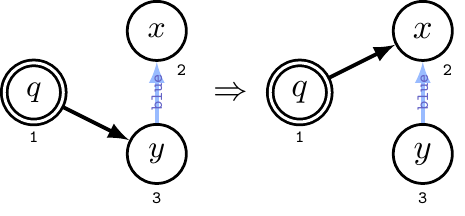}}
  \,:\,  q \in Q,\, x,y \in \Gamma\, \bigg\} $
  
%   \medskip
  $\ttt{HeadRight} = \bigg\{ \,
    \raisebox{-.55\height}{\includegraphics[scale=.9]{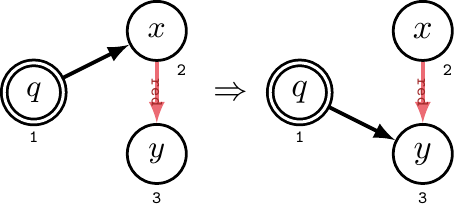}}
  \,:\,  q \in Q,\, x,y \in \Gamma\, \bigg\} $

    \normalsize
    
  \end{tttt}  
  \end{minipage}}
    \caption{Procedures \ttt{NextBlock} and \ttt{PrevBlock} that change the active block.}
    \label{fig:block}
  \end{figure}

%% file: Figures/Encode/encode.tex
\begin{figure}
  \centering
  \fbox{\begin{minipage}{.98\textwidth} 
  \begin{tttt}\scriptsize
    Encode = EncodeInit; Encoding!; Update

    Encoding = CacheDec; try Finish then break; try next\_value

    % Update = \{update1, update2, update3, update4, update5\}

    \smallskip
    CacheDec = CacheInit; Decrement!; if Unfinished then Reset!

    Decrement = try Dec then (Finish; break) 
    
    \phantom{....................}else (underflow; try CacheNext else break)
    
    \smallskip
    Reset = overflow; try CachePrev else break

    \medskip

    $\ttt{EncodeInit} = \bigg\{ \,
  \raisebox{-.55\height}{\includegraphics[scale=.75]{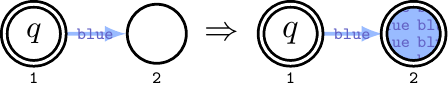}}
  \,:\, q \in Q\, \bigg\} $\quad \quad
      \raisebox{-.40 \height}{\includegraphics[scale=.9]{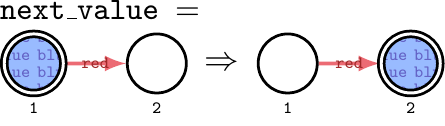}}

    % \smallskip
    $\ttt{Update} = \bigg\{$
    
    \begin{tabular}{ p{48mm} p{48mm} p{45mm} }\vspace{.1pt}
      \includegraphics[scale=.9]{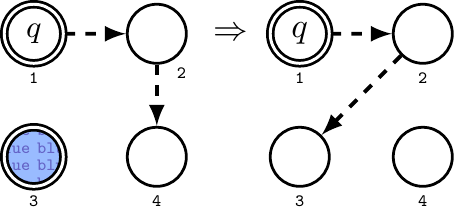}$\quad,$&\vspace{.1pt}
      \includegraphics[scale=.9]{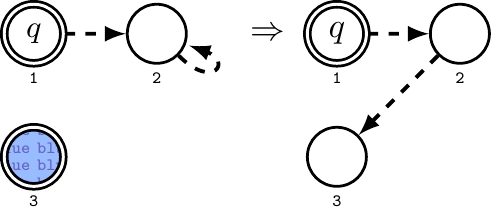}$\quad,$&\vspace{.1pt}
      \includegraphics[scale=.9]{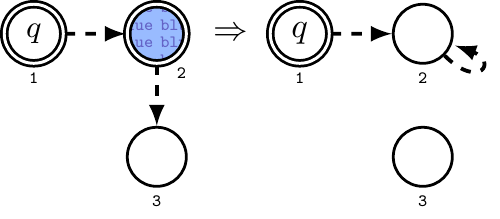}$,$\\\vspace{.1pt}
      \includegraphics[scale=.9]{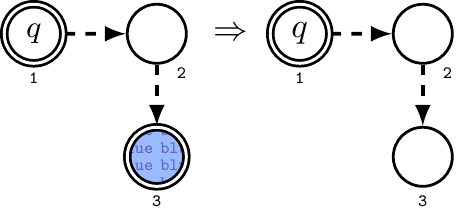}$\quad,$&\vspace{.1pt}
      \raisebox{-.55\height}{\includegraphics[scale=.9]{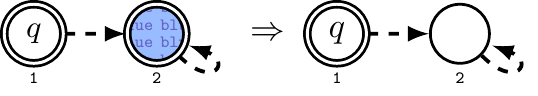}}&\vspace{.1pt} $: q\in Q \,\bigg \}$
    \end{tabular}
    
    % \smallskip
    % \quad\begin{tabular}{ p{45mm} p{60mm} }\vspace{.1pt}
    %   \includegraphics[scale=.75]{Figures/Encode/update1/update1.pdf},&\vspace{.1pt}
    %   \includegraphics[scale=.75]{Figures/Encode/update2/update2.pdf},\\\vspace{.1pt}
    %   \includegraphics[scale=.75]{Figures/Encode/update3/update3.pdf},&\vspace{.1pt}
    %   \includegraphics[scale=.75]{Figures/Encode/update4/update4.pdf},\\\vspace{.1pt}
    %   \raisebox{-.55\height}{\includegraphics[scale=.75]{Figures/Encode/update5/update5.pdf}}&\vspace{.1pt} $: q\in Q \,\bigg \}$
    % \end{tabular}

    \smallskip
    
    $\ttt{CacheInit} = \bigg\{ \,
  \raisebox{-.55\height}{\includegraphics[scale=.9]{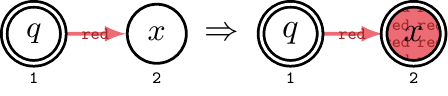}}
  \,:\, q \in Q,\, x \in \Gamma \, \bigg\} $
  \quad
  $\ttt{Dec} = \bigg\{ \,
  \raisebox{-.55\height}{\includegraphics[scale=.9]{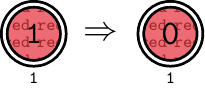}}
  \quad$,$\quad 
  \raisebox{-.55\height}{\includegraphics[scale=.9]{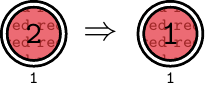}}
  \, \bigg\} $ 
  
%   \smallskip

  % \begin{tabular}{ p{58mm} p{20mm} p{20mm} }
  %   \hspace{-2mm}$\ttt{Dec} = \bigg\{ \,
  % \raisebox{-.55\height}{\includegraphics[scale=.9]{Figures/Count/dec/dec1.pdf}}
  % \quad$,$\quad 
  % \raisebox{-.55\height}{\includegraphics[scale=.9]{Figures/Count/dec/dec2.pdf}}
  % \, \bigg\} $  &
  %   \raisebox{-.40\height}{\includegraphics[scale=.9]{Figures/Count/over/over.pdf}} &
  %   \raisebox{-.40\height}{\includegraphics[scale=.9]{Figures/Count/under/under.pdf}} 
  %   \end{tabular}
    
  %   % \smallskip
  % \begin{tabular}{ p{50mm} p{50mm} }
  %   \hspace{-2mm}$\ttt{Finish} = \bigg\{ \,
  % \raisebox{-.55\height}{\includegraphics[scale=.9]{Figures/Count/finish/finish.pdf}}
  % \,:\, x \in \Gamma \, \bigg\} $ &
  % $\ttt{Unfinished} = \bigg\{ \,
  % \raisebox{-.55\height}{\includegraphics[scale=.9]{Figures/Count/unfinished/unfinished.pdf}}
  % \,:\, x \in \Gamma \, \bigg\} $
  % \end{tabular}

  \raisebox{-.40\height}{\includegraphics[scale=.9]{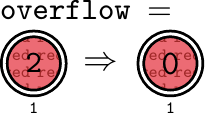}} 
  \quad \quad
    \raisebox{-.40\height}{\includegraphics[scale=.9]{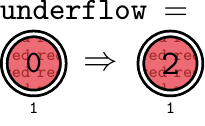}} 
\quad \quad
    $\ttt{Finish} = \bigg\{ \,
  \raisebox{-.55\height}{\includegraphics[scale=.9]{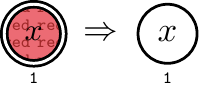}}
  \,:\, x \in \Gamma \, \bigg\} $ 
  \quad
  $\ttt{Unfinished} = \bigg\{ \,
  \raisebox{-.55\height}{\includegraphics[scale=.9]{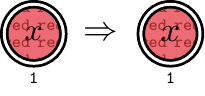}}
  \,:\, x \in \Gamma \, \bigg\} $
  
%   \smallskip
  $\ttt{CacheNext} = \bigg\{ \,
  \raisebox{-.55\height}{\includegraphics[scale=.9]{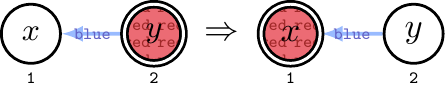}}
  \,:\, x,y \in \Gamma \, \bigg\} $
  \quad
  $\ttt{CachePrev} = \bigg\{ \,
  \raisebox{-.55\height}{\includegraphics[scale=.9]{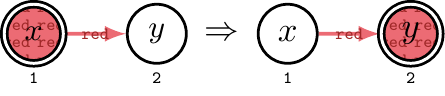}}
  \,:\, x,y \in \Gamma \, \bigg\} $
    
    \normalsize
    
  \end{tttt}  
  \end{minipage}}
    \caption{Procedure \ttt{Encode} that encodes the current block into BLOCKSET.}
    \label{fig:encode}
  \end{figure}

%% file: Figures/Decode/decode.tex
\begin{figure}
  \centering \scriptsize
  \fbox{\begin{minipage}{.98\textwidth} 
  \begin{tttt}
    Decode = DecodeInit; Decoding!; Update

    Decoding = try prev\_value else break; CacheInc

   \smallskip
   CacheInc = CacheInit; Increment!; if Unfinished then (Reset!; Finish)

    Increment = try Inc then (Finish; break)
    
    \phantom{....................}else (overflow; try CacheNext else break)

    \smallskip
    
    $\ttt{DecodeInit} = \bigg\{ \,
  \raisebox{-.55\height}{\includegraphics[scale=.9]{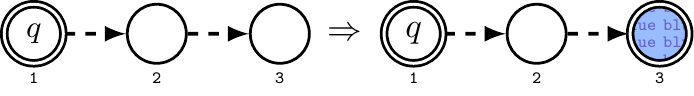}}
  \,:\, q \in Q\, \bigg\} $
  
%   \smallskip
%   $\ttt{Zero} = \bigg\{ \,
%   \raisebox{-.55\height}{\includegraphics{Figures/Decode/zero/zero.pdf}}
%   \,:\, q \in Q\, \bigg\} $

    % \smallskip
    prev\_value $=$ %\begin{tabular}{ p{60mm} p{60mm} }
      \raisebox{-.55\height}{\includegraphics[scale=.9]{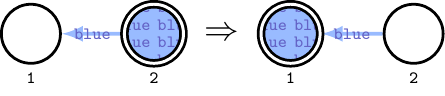}} %&
   \quad \quad
$\ttt{Inc} = \bigg\{ \,
  \raisebox{-.55\height}{\includegraphics[scale=.9]{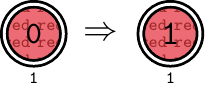}}
  \quad$,$\quad 
  \raisebox{-.55\height}{\includegraphics[scale=.9]{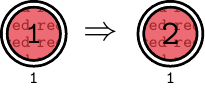}}
  \, \bigg\} $
    
    \normalsize
  \end{tttt}  
  \end{minipage}}
    \caption{Procedure \ttt{Decode} that decodes the current block into CACHE.}
    \label{fig:decode}
  \end{figure}

%% file: Figures/FlagOps/restart.tex
\begin{figure}
  \centering\scriptsize
  \fbox{\begin{minipage}{.98\textwidth} 
  \begin{tttt}
    
    Restart = RewindTapes; ResetCache; ResetBlockset
    
    RewindTapes = try RewindInput; try rewind\_blockset; RewindCache!
    
    ResetCache = CInit; Erase!; end

    ResetBlockset = binit; try Undirect; (copy; try Undirect)!; glue; direct!; unroot
    
    % ResetBlockset = binit; try make\_loop; (copy; direct; try Redirect)!; glue
    
  \end{tttt}
  
  \medskip
  $\ttt{RewindInput} = \bigg\{ \,
    \raisebox{-.52\height}{\includegraphics[scale=.9]{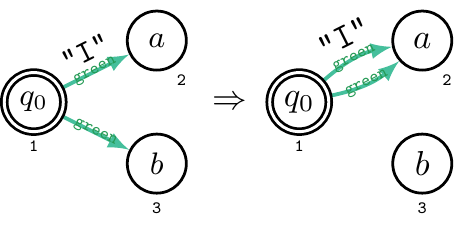}}
  \,:\,   a,b \in \Sigma\, \bigg\} $
  \quad
  $\ttt{rewind\_blockset} \,=\,
    \raisebox{-.55\height}{\includegraphics[scale=.9]{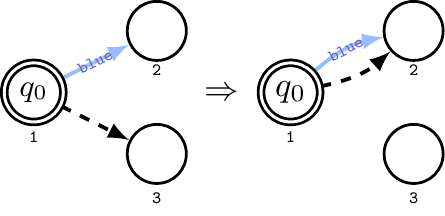}}
   $
  
$\ttt{RewindCache} = \bigg\{ \,
    \raisebox{-.55\height}{\includegraphics[scale=.9]{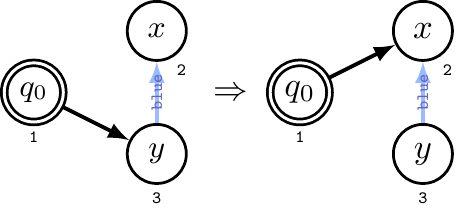}}
  \,:\,  x,y \in \Gamma\, \bigg\} $
  \quad
  $\ttt{CInit} = \bigg\{ \,
    \raisebox{-.55\height}{\includegraphics[scale=.9]{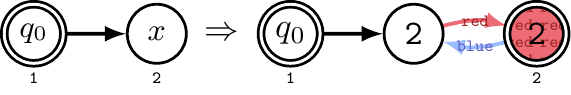}}
  \,:\,  x \in \Gamma\, \bigg\} $
  
  $\ttt{Erase} = \bigg\{ \,
    \raisebox{-.55\height}{\includegraphics[scale=.9]{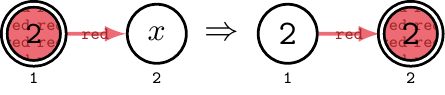}}
  \,:\,  x \in \Gamma\, \bigg\} $
  \quad \quad
  $\ttt{end} \, = \,
  \raisebox{-.55\height}{\includegraphics[scale=.9]{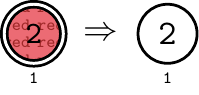}} $
  
  \medskip
  $\ttt{binit} \,=\, 
    \raisebox{-.48\height}{\includegraphics[scale=.9]{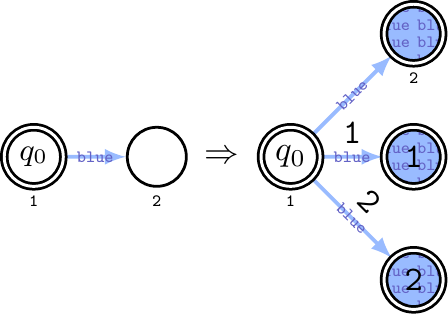}}
  $
  \quad \quad \quad
  $\ttt{copy} \, = \,
  \raisebox{-.48\height}{\includegraphics[scale=.9]{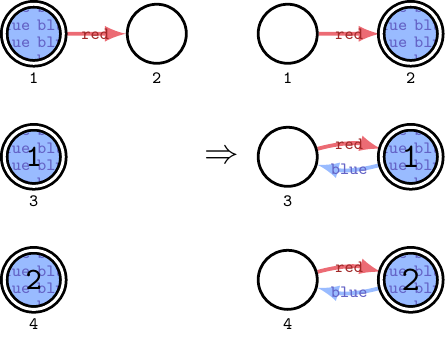}} $

    % \medskip
    $\ttt{Undirect} = \bigg\{ \,
    \raisebox{-.55\height}{\includegraphics[scale=.9]{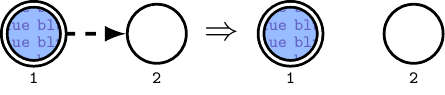}}
  \,\quad,\quad\,  
  \raisebox{-.35\height}{\includegraphics[scale=.9]{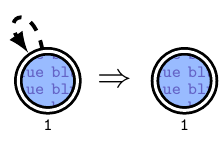}} \, \bigg\}$
  
  % \medskip
  % $\ttt{make\_loop} \, = \,
  % \raisebox{-.55\height}{\includegraphics[scale=.9]{Figures/FlagOps/make-loop.pdf}} $
  % \quad
  % $\ttt{direct} \,=\, 
  %   \raisebox{-.55\height}{\includegraphics[scale=.9]{Figures/FlagOps/direct.pdf}}
  % $
  
  % \smallskip
  $\ttt{glue} \,=\,
    \raisebox{-.47\height}{\includegraphics[scale=.9]{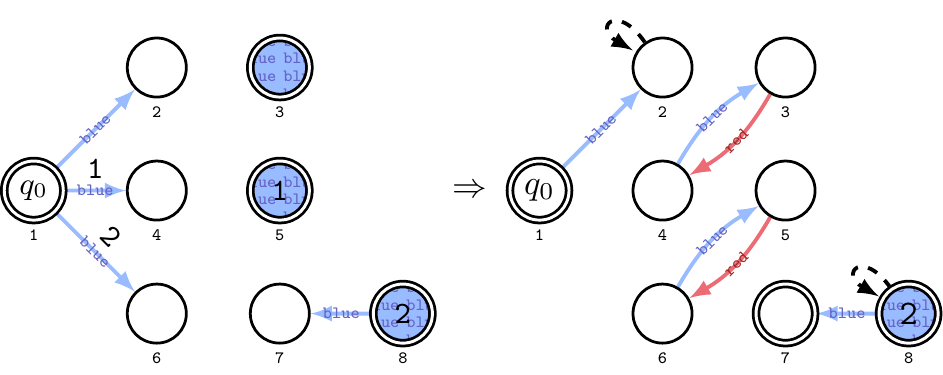}}
  $

  \medskip
  $\ttt{direct} \,=\,
    \raisebox{-.55\height}{\includegraphics[scale=.9]{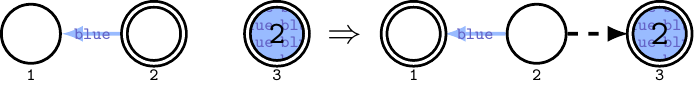}}
  $\quad \quad
  $\ttt{unroot} \, = \,
  \raisebox{-.55\height}{\includegraphics[scale=.9]{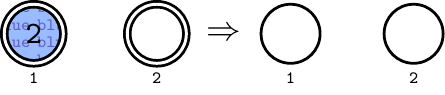}} $

  \normalsize
  
  \end{minipage}}
    \caption{Procedure \ttt{Restart} that resets the simulation and enlarges the tape.}
    \label{fig:restart}
  \end{figure}